\documentclass[5p,times]{elsarticle}

\usepackage{hyperref}
\usepackage{graphicx}
\usepackage{listings}
\usepackage{xcolor}
\usepackage{booktabs}
\usepackage{subcaption}
\usepackage{caption}
\usepackage{balance}
\usepackage{wrapfig}
\usepackage{paralist}

\newcommand\code[1]{{\small \texttt{#1}}}



\usepackage{listings}
\usepackage{siunitx}

\usepackage{enumitem}
\setlist[itemize]{leftmargin=*}

\definecolor{delim}{RGB}{20,105,176}
\definecolor{numb}{RGB}{106, 109, 32}
\definecolor{string}{rgb}{0.64,0.08,0.08}

\lstdefinelanguage{json}{
    frame=tb,
    escapechar=\%,
    postbreak=\raisebox{0ex}[0ex][0ex]{\ensuremath{\color{gray}\hookrightarrow\space}},
    basicstyle=\ttfamily\footnotesize,
    upquote=true,
    morestring=[b]",
    stringstyle=\color{string},
    keywords={from,import},
    literate=
     *{0}{{{\color{numb}0}}}{1}
      {1}{{{\color{numb}1}}}{1}
      {2}{{{\color{numb}2}}}{1}
      {3}{{{\color{numb}3}}}{1}
      {4}{{{\color{numb}4}}}{1}
      {5}{{{\color{numb}5}}}{1}
      {6}{{{\color{numb}6}}}{1}
      {7}{{{\color{numb}7}}}{1}
      {8}{{{\color{numb}8}}}{1}
      {9}{{{\color{numb}9}}}{1}
      {\{}{{{\color{delim}{\{}}}}{1}
      {\}}{{{\color{delim}{\}}}}}{1}
      {[}{{{\color{delim}{[}}}}{1}
      {]}{{{\color{delim}{]}}}}{1},
}

\begin{document}

\begin{frontmatter}

\title{Globus Automation Services:\\Research process automation across the space-time continuum}

\author[anl]{Ryan Chard}
\author[anl,uc]{Jim Pruyne}
\author[uc]{Kurt McKee}
\author[uc]{Josh Bryan}
\author[anl,uc]{Brigitte Raumann}
\author[anl,uc]{Rachana Ananthakrishnan}
\author[anl,uc]{Kyle Chard}
\author[anl,uc]{Ian~T.~Foster}

\affiliation[anl]{organization={Argonne National Laboratory},
             addressline={9700 S Cass Ave},
             city={Lemont},
             postcode={60439},
             state={Illinois},
             country={USA}}
             
\affiliation[uc]{organization={University of Chicago},
             addressline={5730 S Ellis Ave},
             city={Chicago},
             postcode={60615},
             state={Illinois},
             country={USA}}

\begin{abstract}
Research process automation---the reliable, efficient, and reproducible execution of linked sets of actions on scientific instruments, computers, data stores, and other resources---has emerged as an essential element of modern science. 
We report here on new services within the Globus research data management platform that enable the specification of diverse research processes as reusable sets of actions, \emph{flows}, and the execution of such flows in heterogeneous research environments.
To support flows with broad spatial extent (e.g., from scientific instrument to remote data center) and temporal extent (from seconds to weeks), 
these \textit{Globus automation services} feature:
1) cloud hosting for reliable execution of even long-lived flows despite sporadic failures;
2) a simple specification and extensible asynchronous action provider API, for defining and executing a wide variety of actions and flows involving heterogeneous resources;
3) an event-driven execution model for automating execution of flows in response to arbitrary events; and 
4) a rich security model enabling authorization delegation mechanisms for secure execution of long-running actions across distributed resources.
These services permit researchers to outsource and automate the management of a broad range of research tasks to a reliable, scalable, and secure cloud platform. 
We present use cases for Globus automation services, describe their design and implementation, present microbenchmark studies, and review experiences applying the services in a range of applications.
\end{abstract}

\end{frontmatter}

\section*{Keywords}

Research process automation; Globus; High-performance computing; Distributed Computing; Scientific Computing; Cloud computing

\section{Introduction}
\label{sec:introduction}

Consider a materials design application that, over days or weeks, is to perform experimental measurements, computational simulations, and data archiving operations at multiple experimental facilities, computers, and data repositories~\cite{stach2021autonomous,LEONG20223124}.
Or, a long-running synchrotron light source experiment that continuously collects data while periodically retraining, on a remote supercomputer, the AI model used to filter results~\cite{liu2021bridging}, and redeploying retrained models to the experiment site.
Such complex, heterogeneous, long-running, and distributed applications are becoming increasingly common as a result of advances in instrumentation, simulation, and AI methods~\cite{trifan2022intelligent}.

Analysis of such applications reveals a mix of structures and requirements that do not map particularly well to any existing automation technology. 
Like business processes~\cite{bpel,conductor}, they require a user friendly notation, human inputs, and the reliable and secure event-driven execution of sequences of actions, repeatedly and autonomously over extended periods;
however, they also must manipulate large datasets and engage specialized resources.
Like scientific workflows, they often manipulate large datasets and employ high-performance computing (HPC) and concurrency for rapid execution;
however, they must also engage with the physical world.
Like machine learning (ML) workflows, they often must manage dynamically updating data and models~\cite{xin2018developers}; however, they must also engage specialized scientific datasets and resources, and deal with long time scales.  

We describe here an automation approach that integrates and extends several existing technologies to meet what we see as five key requirements of these applications:
\begin{inparaenum}[(1)]
\item
\textit{Reliable execution of long-running flows without local workflow system deployments}.
Leveraging public cloud capabilities and our experiences building and operating Globus transfer services~\cite{chard14efficient}, we employ cloud-hosted, replicated services to ensure that flows and their constituent actions execute reliably without user intervention. 
\item
\textit{Simple, reusable specification of the actions to be performed to meet an application goal}.
We adapt the Amazon States Language~\cite{amazonstateslanguage} as a declarative notation for specifying what we call \textit{flows}, sequences of diverse actions used to meet application needs.
\item
\textit{Event-driven, reactive execution model}.
Leveraging the rich literature on pub-sub systems, we incorporate event capture, filtering, and processing methods to permit data-driven flow triggering and execution.
\item
\textit{Easy integration of new action types}.
Inspired by the extensible Web services architecture, we allow users to incorporate new action types simply by providing a service that implements a RESTful action provider interface. 
\item
\textit{Secure execution of long-running actions on distributed resources}.
Leveraging recent developments in authentication and authorization for distributed systems, we adopt mechanisms provided by the OAuth-based Globus Auth system~\cite{GlobusAuth} for delegation and token renewal.
\end{inparaenum}

\begin{figure*}
    \centering
    \includegraphics[width=0.8\textwidth]{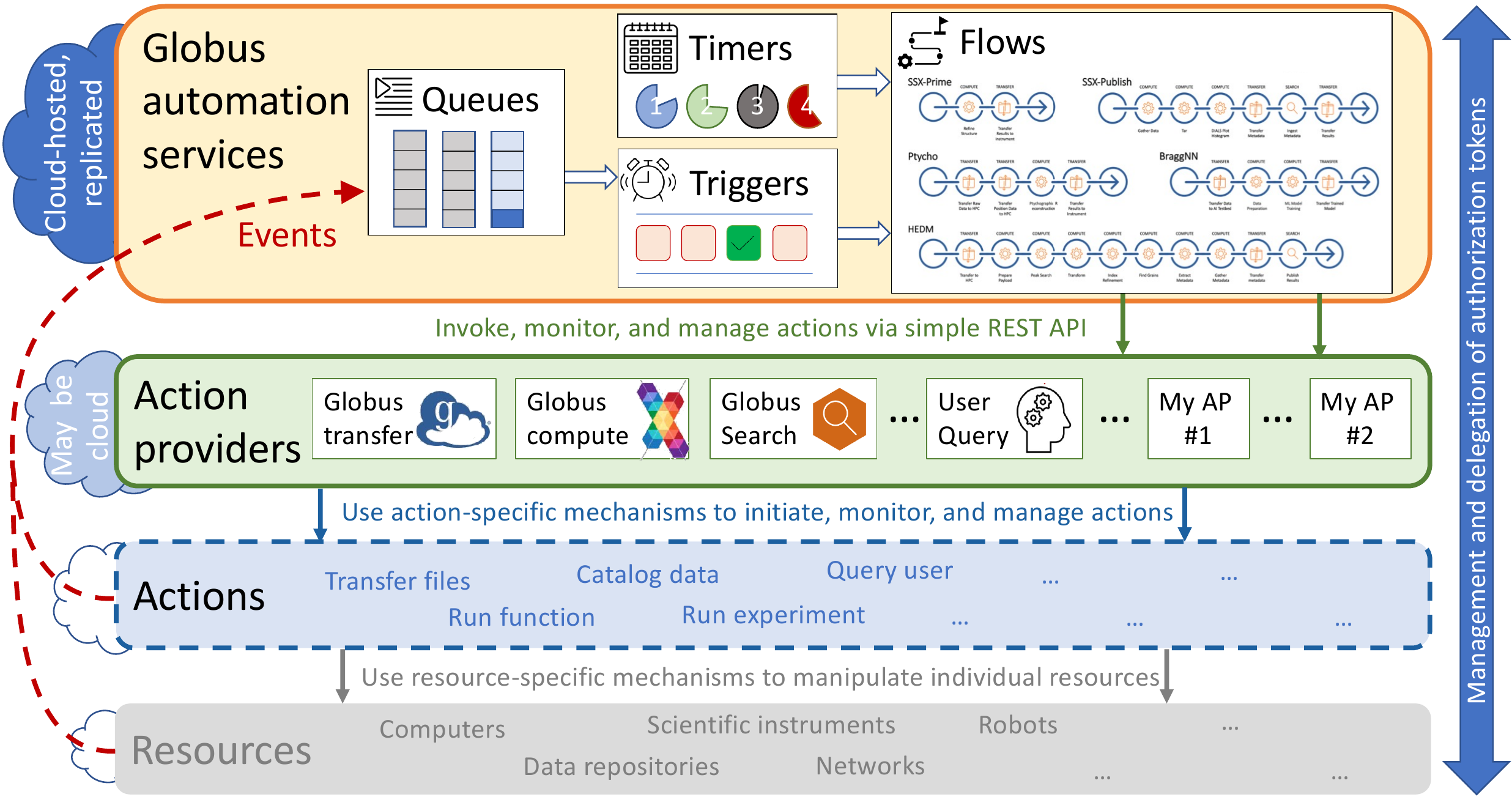}
    \caption{The architecture described in this paper for secure and reliable execution of long-lived, widely distributed research processes. \emph{Automation services} manage the execution of user-specified \emph{flows}. In so doing, they make requests to \emph{action providers}, which in turn initiate and manage \emph{actions} on \emph{resources}. 
    }
    \label{fig:hierarchy}
\end{figure*}

To permit at-scale exploration and evaluation by a diverse community of scientists, we have implemented this new approach as a set of \emph{Globus automation services} (see \autoref{fig:hierarchy}).
These services make it easy to define, for example,  
a \textit{trigger} such that the generation of new data at an instrument causes a \textit{flow} to run, that then engages, in turn, \textit{actions} that 
transfer data to a remote computer,
analyze data, update registries, and email results. 
The action providers that process these actions implement a consistent, asynchronous REST API, facilitating the integration of new activities.

These Globus automation services, like other aspects of the Globus platform~\cite{ananthakrishnan2015globus}, are implemented as platform-as-a-service, i.e., as persistent, cloud-hosted services that are accessible to any authorized user and can be composed to realize different behaviors.
This cloud-hosted approach enables broad delivery 
of research process automation capabilities without requiring that users download and install software, and provides economies of scale that reduce the costs of distributing software.

A diverse community of scientists have been using Globus automation services since 2020 to develop applications that span a wide range of temporal and spatial extents, from the small and local (10$^0$~s tasks on one computer) to the large and distributed (10$^6$~s tasks on computers in distinct authorization domains),
and that encompass diverse numbers, frequencies, and types of actions~\cite{liu2021bridging,vescovi2022linking,bicer2021high,blaiszik2019data,charbonneau2021making,joachimiak2022new,levental2022ultrafast,ali2022fairdms}.

In the remainder of this paper, we present motivating use cases for research process automation, describe Globus automation services and their implementation, present microbenchmark studies of implementation efficiency, and review experiences and lessons learned from early adopters.

\section{Research Process Automation}
\label{sec:motivation}

As research processes in science and engineering become increasingly data-, compute-, and collaboration-intensive, and span ever-larger organizational and geographic regions, there is a likewise increasing need for scalable, reliable, and secure infrastructure to enable their automation. 
This infrastructure must be able to capture multi-step research processes that may span diverse resources (e.g., from instruments to computers; from data centers to the edge) and institutions, and encompass low-latency steering feedback, long-running experiments, and even multi-month data embargoes.

\subsection{Use Cases}
\label{sec:usecases}

We present four use cases in which we have found research process automation to be highly useful. 
From these and other related use cases we distill a set of 
requirements that motivate the design of Globus automation services.
We describe some of these use cases in more detail elsewhere~\cite{vescovi2022linking}.

\subsubsection{Real-time data analysis}

Instruments such as scanning electron microscopes, synchrotron light 
source beamlines, and robotic laboratories can generate large amounts 
of data that must be analyzed, reviewed, catalogued, and shared in a timely manner. 
When analysis requirements exceed storage or processing capabilities co-located with an apparatus, research process automation is needed to move and process data on
more powerful, available, or suitable resources, while navigating security at those resources.

For example, in serial synchrotron crystallography (SSX) experiments at Argonne National Laboratory's Advanced Photon Source (APS), a bright synchrotron beam is used to collect diffraction data from many crystals, at rates of \num{10000}s of images per hour~\cite{diederichs2017serial}.
Experiments typically generate approximately \num{40000} 1475$\times$1255 16-bit pixel images per sample, with tens of samples processed during a beamtime. The experiment is typically configured to generate data at 10Hz (37 MB/s), although much higher rates are possible and will soon become commonplace. 
Images are processed as they are produced, first by the Diffraction Integration for Advanced Light Sources (DIALS) package~\cite{dials} to identify at most one \textit{hit} per image and then, after a number of hits have been identified, with the
post-refinement and merging (PRIME) package~\cite{uervirojnangkoorn2015enabling} to solve the crystal structure.

The SSX processing can be represented as two flows. The first uses seven steps to process each raw image:
\begin{inparaenum}[(1)]
\item Transfer image data from the APS to a high-performance computing (HPC) facility;
\item Perform DIALS Stills processing on each raw image;
\item Extract metadata from files regarding hits;
\item Generate visualizations showing the sample and hit location;
\item Transfer metadata and visualizations for publication;
\item Ingest results, metadata, and visualizations to an SSX Search catalog; and
\item Transfer the results back to the APS.
\end{inparaenum}

After a number of hits, a second flow is run to solve the crystal structure:
\begin{inparaenum}[(1)]
\item Perform PRIME analysis to solve the structure; and
\item Copy the structure back to the APS.
\end{inparaenum}

Similar needs for real-time analysis flows arise in other synchrotron light source experimental modalities (e.g., tomography~\cite{hidayetoglu2021memxct,liu2020tomogan}, x-ray photon correlation spectroscopy~\cite{lehmkuhler2021femtoseconds}, ptychography~\cite{maiden2011superresolution,bicer2021high}, high energy diffraction microscopy~\cite{pokharel18hedm}) and in many other experimental sciences, from cryogenic electron microscopy~\cite{dubochet2012cryo} to multi-messenger astronomy~\cite{huerta2019enabling}. 

\subsubsection{Machine learning training and inference}
ML methods are used increasingly in science for rapid analysis, often near a data source such as for real-time experiment steering. ML models are often refined over time, with model performance improved progressively by training with new and more diverse data as an experiment proceeds.

Here we consider an application of ML methods in high energy diffraction microscopy, a non-destructive technique that combines imaging and crystallography algorithms to characterize polycrystalline materials microstructure in three dimensions (3D) and under various in-situ 
conditions~\cite{bernier2011far,pokharel18hedm}. The technique maps grains in a polycrystalline aggregate by considering diffraction patterns as a function of rotation angle from a synchrotron beam.  

To accelerate the process of  identifying ``spots'' for each microstructure granule, researchers have developed a neural network approximator to identify peak shapes in observed intensities in area detector data. The model is trained with experimental data analyzed with the MIDAS software package~\cite{MIDAS}. Training on a HPC system generates a model that is then deployed on a lightweight device at the instrument for real-time diffraction peak analysis---to enable, for example, experiment steering and anomaly detection. This flow involves four steps:
\begin{inparaenum}[(1)]
    \item Transfer data from experiment to compute facility;
    \item Process data with MIDAS analysis software;
    \item Use many raw/processed data pairs to train a machine learning model using HPC resources and accelerators; and
    \item Transfer the trained model to the edge for inference.
\end{inparaenum}

\subsubsection{Data publication}
\label{sec:datapublication}
Cataloging of data in a registry 
that can be searched both programmatically and via a web interface is essential to
making data findable, accessible, interoperable, and reusable (FAIR). The publication processes used to populate registries typically require the
orchestration of numerous steps over varying time-scales including waiting for human input. 

For example, data publication in to the Materials Data Facility (MDF)~\cite{blaiszik16mdf}
encompasses initial data upload followed by quality control, metadata extraction, curator approval, and metadata indexing in to a catalog. 
MDF relies on a cloud-hosted search index to catalog dataset metadata, and a large storage system for storing the datasets. 
As the publication flow proceeds, some steps must be performed with credentials for the user publishing the data (e.g., moving data to which only the user has access) while others require administrator or system credentials (e.g., assigning an identifier owned by the system).
Similar sequences of actions arise when publishing machine learning models in the Data and Learning Hub for science~\cite{chard19dlhub}, bioinformatics datasets in the Common Fund Data Ecosystem \cite{cfde}, and other datasets at the Argonne Leadership Computing Facility's (ALCF) Community Data Co-op (ACDC)~\cite{allcock2019petrel}.

In the MDF case, the publication flow proceeds as follows. Users start the flow via a web-based portal. The flow then proceeds to:
\begin{inparaenum}[(1)]
    \item Allocate storage for the user to upload data (shared only with the submitter);
    \item Transfer the data from the user's source location to the allocated storage;
    \item Request the submitter to input metadata via a web form;
    \item Apply automated metadata extraction methods to derive metadata from common formats;
    \item Request that a curator either approve data and metadata or return them to the submitter for modification;
    \item Assign a persistent identifier (a DOI from DataCite);
    \item Index metadata in a search index; and
    \item Set final access permissions on data based on system polices and user specification.
\end{inparaenum}

In each of these described cases, while the invocation and management of the various steps could be performed manually, or implemented in a custom script, a managed automation service allows such process flows to be automated at scale and with monitoring and guaranteed progress (i.e., resistance to failure at the location running the script) over an extended time frame.

\subsubsection{Analysis as a service}
Modern scientific simulation, analysis, and learning methods are transforming entire 
science disciplines; however, they are also broadening
the gap between those with access to large-scale and specialized computing resources
and those without. Thus, researchers and computing 
facilities are developing systems that democratize
access to cutting-edge computational capabilities
via accessible, scalable, and easy-to-use interfaces. 
Implementing such services requires a number
of steps including data upload, model execution, 
and notification of results. 

One example is AlphaFold~\cite{senior2020alphafold}, a
deep learning system that predicts protein structures. 
AlphaFold is computationally expensive, requiring GPUs to process sequences in a timely manner.
Researchers at ALCF have developed a service
that enables execution of AlphaFold on demand. Its implementation requires the staged orchestration of several steps.
First, a user employs a web form to 
request inference on an uploaded dataset. 
Subsequent steps then:
\begin{inparaenum}[(1)]
    \item Create a writable path on a shared storage system for the user to upload their data;
    \item Stage the data to an available compute cluster;
    \item Execute AlphaFold on the uploaded data;
    \item Transfer results to a publicly accessible HTTPS server; and
    \item Send an email to the user notifying them that the computation is complete.
\end{inparaenum}

\subsection{Requirements}
Based on these use cases, we identified the following requirements common to research automation scenarios. 

\begin{itemize}
\item 
\textbf{Diverse actions}: Flows may include computational (e.g., analysis, data movement, metadata extraction, persistent identifier minting, indexing, model training), physical (e.g., initiating an experiment), and human (e.g., providing metadata, data curation) actions. 

\item
\textbf{Secure delegation}: Flows may span administrative domains and institutions. Fine-grained and delegatable authorization is required to ensure that actions are performed only when and where authorized.

\item 
\textbf{Automated}: It is often desirable that flows be started without human intervention, for example when data are acquired from an instrument. 

\item 
\textbf{Programmable}: Flows require control logic, such as conditionals to modify behavior based on action results and errors.

\item 
\textbf{Reusable}: 
It is important to permit specification, sharing, and reuse of flow ``recipes,'' so that a flow can be invoked many times, potentially by many people, including those 
not involved in creating or authoring the flow.

\item 
\textbf{Interrogable}: Users must be able to review execution of a flow to understand what, where, and why actions were performed, whether the flow completed successfully, and under what conditions the flow was invoked. 

\item
\textbf{Intuitive}: Flows may be defined and invoked by a range of users (e.g., scientists, students) who require 
intuitive interfaces for defining, invoking, and managing flows.

\item
\textbf{Span time scales}: Research processes, and individual actions, may execute over varying time scales, from second to months.  There is also a need to manage both synchronous and asynchronous actions.

\item 
\textbf{Robust and available}: The platform must enable users to outsource flow management without repeated interactions, and to author flows that can compensate for failed actions. 

\end{itemize}

\section{Globus Automation Services}
\label{sec:design}

The Globus platform~\cite{ananthakrishnan2015globus} comprises an integrated set of services that together provide a consistent view (from an API perspective) of diverse identity and access management (IAM) methods and data and compute resources:
\begin{itemize}
    \item 
        \textbf{IAM services} (Auth~\cite{GlobusAuth}, Groups~\cite{chard16nexus}) for single sign-on, management of identities and credentials, and delegation.
    \item 
        \textbf{Data services} (Transfer~\cite{allen12software}, HTTPS,
        Share~\cite{chard14efficient}) for access to, and managed movement of, data.
    \item
        \textbf{Metadata management} (Search~\cite{ananthakrishnan18platform}, Identifiers~\cite{ananthakrishnan20identifiers}) for indexing and generating persistent references to data.
    \item 
        \textbf{Compute services} (funcX~\cite{chard2020funcx,li2022funcx}, OAuthSSH~\cite{alt20oauthssh}) 
        for invocation and management of computational tasks.  
\end{itemize}

Here we describe four new \textbf{Globus automation services}---Flows, Triggers, Queues, and Timers---that extend the scope of the Globus platform.
Their purpose is to \emph{simplify the definition, deployment, invocation, and management of robust, secure, long-lived, multi-functional research automation flows}.

\autoref{fig:hierarchy} illustrates important elements of these services. 
A flow is organized as a sequence of \textbf{states} that are processed in sequence, with support for conditional execution.
A state can be implemented via an \emph{action provider}, of which we show several in the figure (Transfer, Compute, Search, Query, etc.) or by built-in methods (e.g., Choice).
Action providers often interact with persistent services to perform a requested action: in the figure, those services are data transfer, computation, user query, and publication.

Flows may be invoked manually or automatically as the result of \emph{triggers} or \emph{timers}. 
\emph{Triggers} process \emph{events}, which may be generated remotely and passed via reliable message \emph{queues}.
\emph{Timers} allow for periodic scheduling of flows.  

\textbf{Actions}:
Any activity with some notion of completion can be made accessible as an
\textit{action}.
For example, an action may transfer data between two locations, request human review of a sample, or actuate a robot to place a sample in a microscope. Actions are typically asynchronous, in that a request to start an action returns not a result but an identifier that can be subsequently used to check for success (or failure) and to access any results (or error messages). An action may require input arguments (e.g., source and destination for a data transfer; sample to be reviewed and identity of reviewer; robot movement parameters) to complete its task, and may return purpose-specific information (e.g., transfer progress; review result; robot status).

Globus automation services represent actions as web services that implement the \textit{action provider API}, which defines methods for asynchronous invocation and status monitoring. 
Globus provides several action providers implementing this API, including data transfer, remote computation, human feedback via a web form, notification via email, and minting of persistent identifiers~\cite{globusactionproviderlist}.
The action provider API is open and designed to be implemented by external services; developers can easily create new action providers, either by extending an existing service implementation or by wrapping existing functionality behind the action provider API.

\textbf{Flows}:
A flow defines a sequence of action invocations and other processing steps (e.g., manipulation of flow run Context) and control logic (e.g., Choice, Wait, Fail).
Flows are defined in a declarative manner by specifying individual actions, control logic, and conditions upon which the flow should proceed or halt.
Once published, a flow may be invoked one or many times, by the author or by others authorized to invoke the flow. 
Flows themselves are also action providers and thus can be 
included in other flows. 

Flows maintain a run \textit{Context} throughout their execution which is accessible to and modifiable by actions.
The Context allows the flow and its actions to modify behavior based on the results of previous actions. For example, a transfer action can set a filename for subsequent use by an analysis action.

\textbf{Events and queues}:
Globus automation services support the automated invocation of flows in response to a variety of events. 
As with actions, we take a broad view of potentially interesting events that may encompass, for example,
events generated by actions or flows, 
file systems or instruments, or humans (e.g., via email or web pages). 

Given the wide range of events that we may wish to have spur actions, and the fact that these events may occur in different places at different times, we require an extensible, yet common abstraction that decouples event generation from action execution. This decoupling includes both time and location, but also identity: the source of the event and the consumer of the event may have different Globus identities. For this purpose, we adopt a queue-based model as an intermediary between events and actions on those events. 
Users define specific Globus automation service queues and may deploy remote event
generators (e.g., on local filesystem) to send events to these queues. 

\textbf{Triggers}:
As event generation and consumption are loosely coupled, we want to be able to filter event streams to focus on events of interest.
Globus automation services use triggers as an event-independent
way of responding to events with specified characteristics. A trigger
defines an event source, a predicate on the content of events, and the action(s) to perform when the predicate is satisfied. (At the time of writing, the Triggers service is a prototype.)

\textbf{Timers} are provided by Globus automation services
to enable actions to be invoked periodically,
at specified intervals.

\section{Using Globus Automation Services}

We describe here how users interact with Globus automation services, to provide context for the implementation descriptions that follow. 

\subsection{Interfaces and tools}

\begin{figure*}
    \centering
    \begin{subfigure}[b]{0.8\textwidth}
         \setlength{\fboxsep}{0pt}\fbox{\includegraphics[width=\textwidth,trim=0 36mm 0 103,clip]{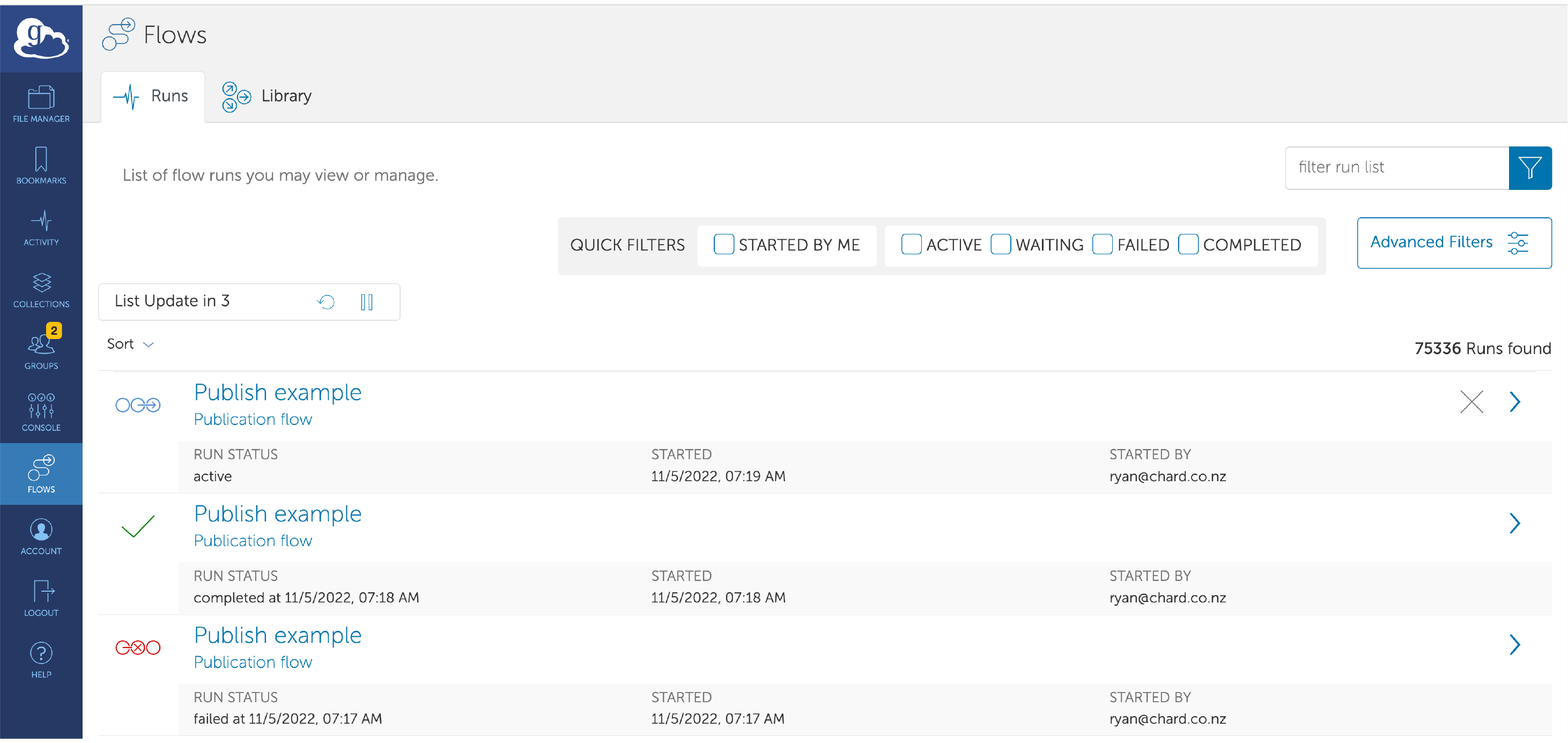}}
        \caption{The \textbf{Runs} tab in the Flows interface lists runs that I can view or manage. (The \textbf{Library} tab lists flows that I can run.)}
        \label{fig:XPCS_flows_a}
    \end{subfigure}
    
    \vspace{2mm}

    \begin{subfigure}[b]{0.8\textwidth}
          \setlength{\fboxsep}{0pt}\fbox{\includegraphics[width=\textwidth,trim=0 40mm 0 113,clip]{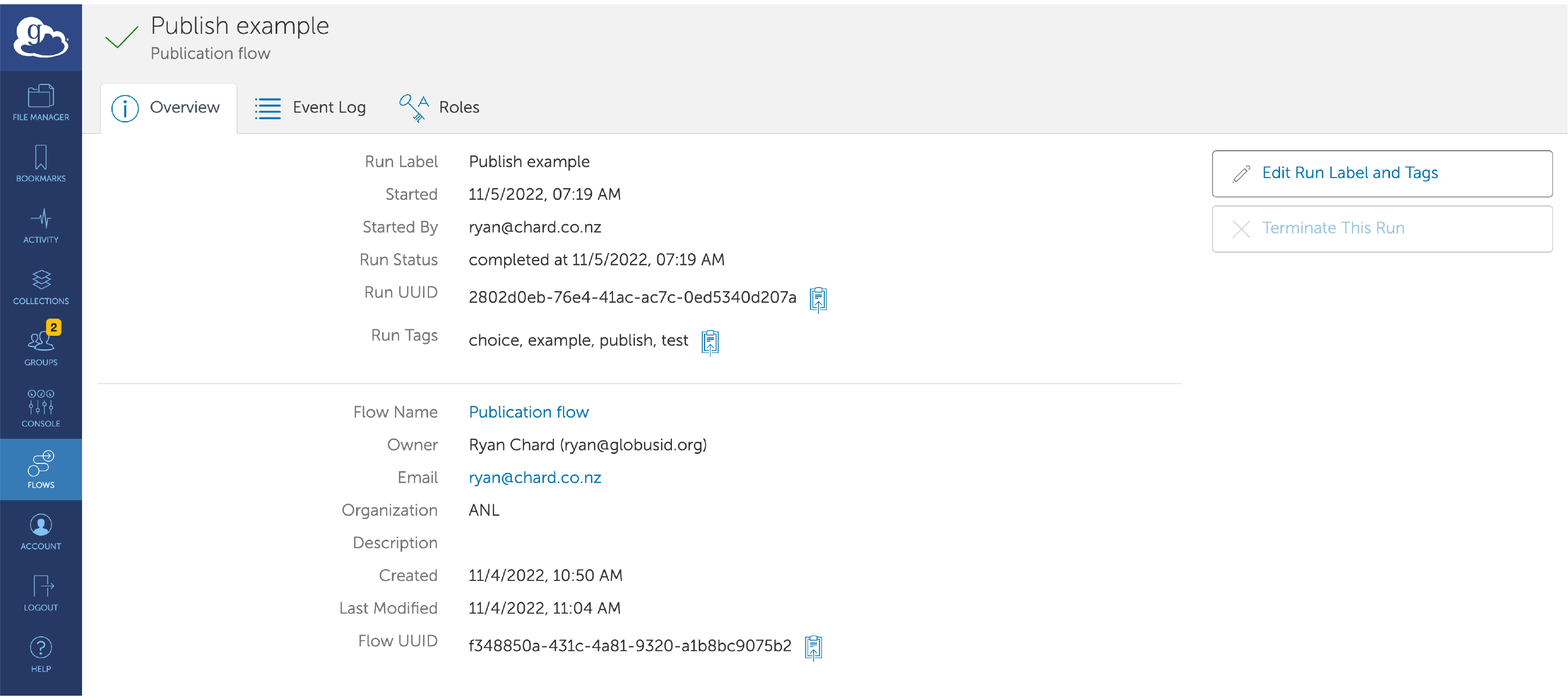}}
        \caption{Selecting a run in \autoref{fig:XPCS_flows_a} gives information on the run (above) and the flow that was run (below).}
        \label{fig:XPCS_flows_b}
    \end{subfigure}
    
    \vspace{2mm}
    
    \begin{subfigure}[b]{0.8\textwidth}
        \setlength{\fboxsep}{0pt}\fbox{\includegraphics[width=\textwidth,trim=0 38mm 0 106,clip]{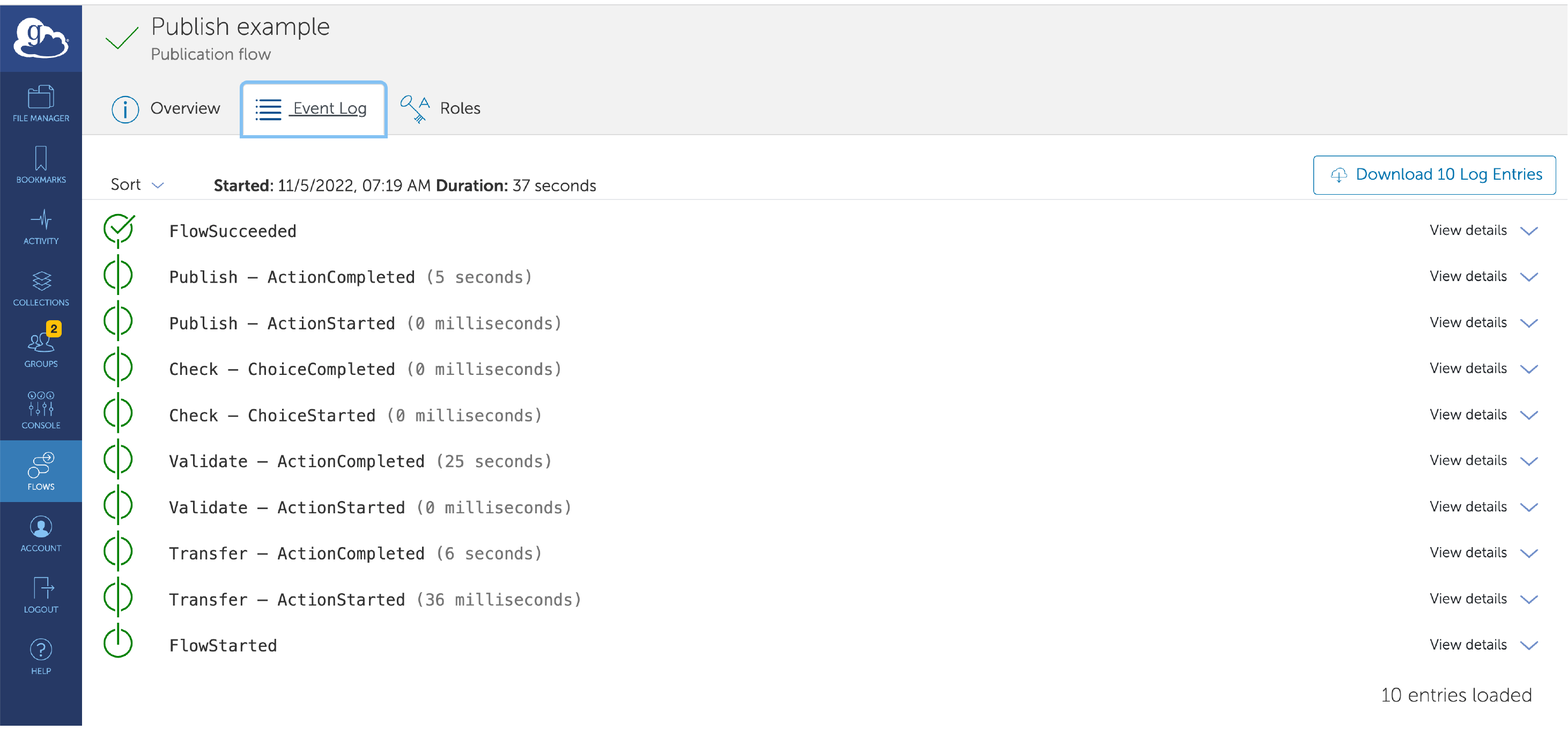}}
        \caption{The \textbf{Event Log} tab in \autoref{fig:XPCS_flows_b} shows a (here truncated) log of flow processing events during the run. All actions completed successfully.}
        \label{fig:XPCS_flows_c}
    \end{subfigure}
    
    \caption{We use a simple publication flow to illustrate how the Globus web interface enables tracking of flow progress and diagnosing of errors by (a) listing recent runs, (b) inspecting a summary of a run, and (c) listing actions involved in that run (shown with most recent first). Other displays, not shown here, allow for examination of flow definitions, input schema, and actions in failed runs.
    }
    \label{fig:XPCS_flows}
\end{figure*}

Globus automation services support three client interfaces for interacting with the services: 
a Python SDK, which implements a client class for programmatic invocations;
a command line interface (CLI), for interactive or scripting use;
and, for general use, a web application to run and manage flows.
The web application facilitates not only
running flows but also detecting, diagnosing, and correcting errors that may occur when a flow is executing, as shown in \autoref{fig:XPCS_flows}.

\subsection{Working with Flows}

Globus automation services allow users to \textit{author}, \textit{publish}, \textit{discover}, \textit{invoke}, and \textit{manage} flows. 

Authorized users can \textbf{author} a flow by creating 1) a \emph{definition}, which specifies the set of states comprising the flow;
2) an \emph{input schema}, specifying constraints on input data to a flow run; 3) \emph{permissions} governing visibility, use (i.e., ability to run) and management of a flow and; 4) \emph{metadata}, including a title, description, and searchable keywords.

Having authored a flow, the user can \textbf{publish} it to the Flows service. The service validates the flow definition and input schema, and deploys the flow (i.e., makes it available for execution) and returns a unique flow id. 
Flows serves as an action provider factory, creating a new API path (\code{flows/<flow\_id>}) supporting the action provider API for operations on the flow.
Thus, any authorized person (or program) can then employ the action provider API (see \autoref{sec:APs}), with the constructed API path, to introspect or invoke the flow. One such use of the API may be the Flows service itself allowing a ``parent`` flow to specify a different, ``child,'' flow as an action state.

Any authorized user can then search or browse to \textbf{discover} flow(s) with desired characteristics.
Having identified a suitable flow, a user may 
\textbf{invoke} it to create what we call a \emph{run}, supplying values to be populated into the run's Context and which must satisfy the input schema.
Once a run is created, a user can \textbf{manage} it: \emph{monitor} its status, \emph{terminate} it prior to completion, and/or \emph{retrieve} either results upon successful completion or error reports upon failure.

Each step after authoring can be performed via the CLI or the SDK. The web application can be used for all operations except authoring and publishing.

\subsubsection{Flow definition}

We use a declarative language to author flow definitions. This language extends
the Amazon States Language (ASL)~\cite{amazonstateslanguage} used to define Amazon Step Functions~\cite{amazonstepfunctions} state machines in the AWS cloud. 
The flow definition's JSON format is rather verbose, so we focus here on describing its primary features rather than detailed syntax.

A flow definition specifies, first of all, a set of state definitions 
plus the start state. 
For example, the following code fragment describes a flow with five states (Transfer, Validate, Check, Publish, Failure) that is to start in the Transfer state.

\begin{lstlisting}[language=json]
{ "StartAt": "Transfer",
  "States" : {
    { "Transfer" :
       { "Type":"Action", ...,
         "Next":"Validate" }},
    { "Validate" : 
       { "Type":"Action", ..., 
         "Next":"Check" }},
    { "Check" : 
       { "Type":"Choice", ...,
         <"Failure" or "Publish">}},
    { "Publish" :
       { "Type":"Action", ..., 
         "End": true}},
    { "Failure":
       {"Type": "Fail" }, ... }}
}
\end{lstlisting}

Flows supports five distinct types of state: four taken essentially unchanged from ASL (Choice, Pass, Fail, Wait), plus an Action state, used to invoke an action provider.
The Transfer, Validate, and Publish states used in the example to perform data transfer, data validation, and data publication actions have type Action, while Check, which is used to check the result of Validate to see whether the input data should be published, has type Choice. Finally, the Failure state is of type Fail, which causes the run to terminate and to register as an abnormal exit.

Each state type requires additional information.
For example, an Action will always specify a URL for the associated action provider and will typically also provide 
information about input and output values and a timeout value.
These are seen in the following skeleton for the Validate action, which uses the funcX~\cite{chard2020funcx,li2022funcx} action provider to run a function \texttt{validate}. The skeleton specifies parameters (including the input data: the payload) passed as a \texttt{tasks} parameter and the output result returned at \texttt{Valid}. 
The prefix \texttt{\$.} on these values signals that they should be treated as JSONPath references into the run Context. The \texttt{WaitTime} indicates that the instance should wait no longer than 7200 seconds (two hours) for the action to complete, after which it should treat the action as a failed state.

\begin{lstlisting}[language=json]
"Validate": {
  "Type": "Action",
  "ActionUrl": "https://automate.funcx.org/",
  "Parameters": {
    "tasks": [
      {
        "endpoint.$": "$.endpoint_compute",
        "function.$": "$.validate_function_id",
        "payload.$": "$.payload"
      }
    ]
  },
  "ResultPath": "$.Valid",
  "WaitTime": 7200,
  "Next": "Check"
}
\end{lstlisting}   
   
The ability to catch and respond appropriately to failures is essential to any process automation system. Flows, like the ASL that it extends, allows the author to specify alternate control flow upon failure, such as in the following which, if added to the Validate state definition above would cause the flow to transition to the Failure state upon failure with error information returned into the run Context under the key \texttt{ValidFailureInfo}.

\begin{lstlisting}[language=json]
  "ExceptionOnActionFailure": true,
  "Catch": [{
    "ErrorEquals": 
      [ "ActionFailedException" ],
      "ResultPath": "$.ValidFailureInfo",
      "Next": "Failure"
    }]
\end{lstlisting}

As we discuss in more detail in \autoref{sec:auth}, Globus automation services permit fine-grain control over the identity used to perform different actions. 
By default, actions are run as the run creator (the user who invoked the flow), but flow authors can also specify alternatives. 
For example, adding the following to the Validate state definition specifies that the action should run as \texttt{ComputeProvider}:

\begin{lstlisting}[language=json]
  "RunAs": "ComputeProvider"
\end{lstlisting}

When the flow is invoked, this role is mapped to the
identity under which the validation computation should occur, and credentials (in the form of Globus Auth generated OAuth tokens) for that identity are provided when invoking the flow.
This pattern is useful for end-user facing services that use Globus automation services behind a portal or other interface (such as the use case described in \autoref{sec:datapublication}) in which some flow states need to access resources, like datasets, for which an end user needs to grant permission, while other action states require other credentials (e.g., system credentials to provision resources). It is also useful in the case of service-owned resources such as an HPC system, for which a distinct identity (e.g., that of a group account) is required. 

\subsubsection{Flow execution model}
Each run of a flow has associated with it a \emph{Context}, which takes the form of a JSON document. This Context is initialized with the input values provided when the flow is invoked. Each state in a flow may read or write values to/from this Context, with the location within the Context specified using JSONPath syntax. Upon completion of a flow, whether successful or not, the final Context is returned to the user (or to other flows, triggers or timers which may have invoked the flow).

\subsubsection{Flow Input Schema}

\begin{figure}[ht]
  \vspace{-0in}
  \centering
  \fbox{\includegraphics[width=0.6\columnwidth]{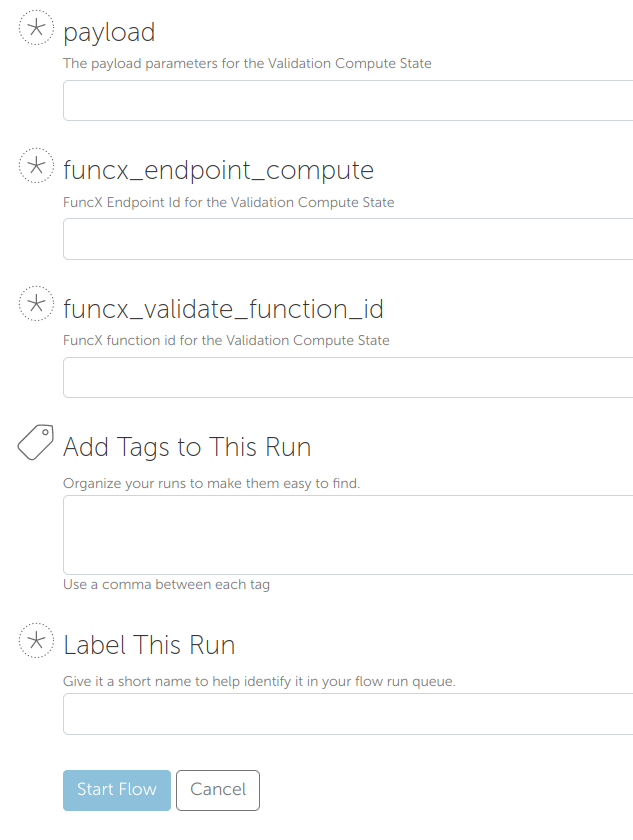}}
  \vspace{-2mm}
  \caption{Browser-based input form generated from a flow's input schema. This example reflects the required input of a simple flow that uses the Validate action shown above, 
requiring the user to specify a payload, funcX endpoint id, function id, optional tags, and an optional label for the run. 
  }
\label{fig:input-form}
  \vspace{-0.1in}
\end{figure}

Each flow must include a schema, defined in JSON Schema \cite{bhutton-json-schema-00} syntax, for validating input when running the flow. Validation via the schema prior to running a flow makes run-time failure due to improper input less likely. 
The input schema is also useful for building user interfaces and clients for starting runs. 
For example, the web application uses the input schema to dynamically render a form 
(see \autoref{fig:input-form}) 
to guide a user in providing required flow input.

\subsection{Access control}
\label{sec:flow-access-control}
All use of Globus automation services 
is subject to authorization, which requires authentication and may also be subject to group membership or other access control requirements.

Role-based access control is used to mediate who can perform various actions on flows and runs. Permissions may be granted to an individual user identity, a group, or an application identity in the Globus ecosystem. 
The following roles are supported on a flow:
\begin{itemize}
    \item \textbf{Visible To}: May discover and display, but not run, the flow. The special value \textit{public} can be used to indicate that a flow is visible without authentication and thus may be viewed by any user.
    \item \textbf{Runnable By}: May invoke the flow.  The special value \textit{all authenticated users} indicates that any user who presents valid credentials
    may run the flow.
    \item \textbf{Administered By}: May update the flow, including changes to the definition, schema, and descriptive metadata.
    \item \textbf{Owner}: The identity of the user who initially published the flow. Only the Owner may remove the flow. This value may be re-assigned by an administrator, for example, to account for an owner leaving the organization that maintains a flow.
\end{itemize}

The permissions associated with these roles are cumulative: for example, an identity in \textit{Runnable By} is inherently also in \textit{Visible To}.

To facilitate management of runs, two additional roles are supported: \emph{Monitor} with permissions to view a run including its initial input, its progress and its final result, and \emph{Manager} to view (cumulative with the \textit{Monitor} role) and cancel a run.

\subsection{Defining Triggers and Timers}
Triggers and timers provide two mechanisms for event-driven flow execution. Users may define and deploy a trigger or timer to automatically invoke a flow when an event condition is met or on a schedule.

A trigger definition specifies:
\begin{inparaenum}[(1)]
    \item a \textbf{queue identifier} for the persistent queue from which events are to be received; 
    \item a \textbf{predicate}, expressed as a filter on event parameters with Boolean result, under which a flow should be invoked;
    \item the \textbf{flow} (or action) to be invoked if the predicate is satisfied; and  
    \item a JSON \textbf{template} of the body that will be passed to the flow (or action) when run.
\end{inparaenum}

A timer definition specifies: 
\begin{inparaenum}[(1)]
    \item a \textbf{schedule} (i.e., start time, interval) for invoking a flow (or action);
    \item the \textbf{flow} (or action) to be invoked;
    \item a JSON \textbf{template} of the body that will be passed to the flow (or action) when run.
\end{inparaenum}
(A timer may be viewed as a specialized form of trigger.)

\subsection{Defining queues}
Queues is a managed service that allows users to provision queues on demand. Queues may be created via the SDK and CLI. Users can specify configuration options, such as message timeout duration and access roles. Once provisioned, authorized users may publish arbitrary events (as a JSON payload) to the queue using the SDK and CLI. Events can therefore be generated from external services, scripts, and file system monitoring software, for example. Events can be consumed from a queue using the SDK or CLI, or with a Trigger.

\subsection{Example Actions}
\label{sec:actionsusage}
We evaluate the following seven action providers in \autoref{sec:evaluation}. A complete list of available action providers is provided in the documentation~\cite{globusactionproviderlist}.

\begin{itemize}

    \item \textbf{Echo}: Returns its input string, and is primarily used for testing and demonstration.
    
    \item \textbf{Transfer}: 
    List directories, manage permissions, delete data, transfer data between remote systems. 
  
    \item \textbf{Search}: Add/delete entries to/from a search index.
    
    \item \textbf{Email}: Send a templated email with specified 
    sender, receiver(s), subject, and body. 
    Templates allow values from the flow run Context to be included in the body of the email.
    
    \item \textbf{User Selection}: An interactive action that enables users to provide feedback via
    a list of options or customized interface;
    user selection(s) are returned to the flow. 
    
    \item \textbf{GenerateDOI}: Obtain a DataCite DOI to assign to a web-accessible object. The action provider uses the DataCite JSON API and allows users to preconfigure it with the appropriate namespace and DataCite basic auth credentials. Invocation of the action passes through JSON metadata to be associated with the DOI.

    \item 
    \textbf{funcX}:
    Request execution of a registered Python function on a remote computer. Users specify the funcX endpoint ID and function ID as well as any input arguments. 
    The action provider wraps calls to the funcX~\cite{chard2020funcx,li2022funcx} function as a service (FaaS) platform.

\end{itemize}

\section{Implementation}
\label{sec:implementation}

\begin{figure*}[th]
  \vspace{-0in}
  \centering
  \includegraphics[width=0.9\textwidth]{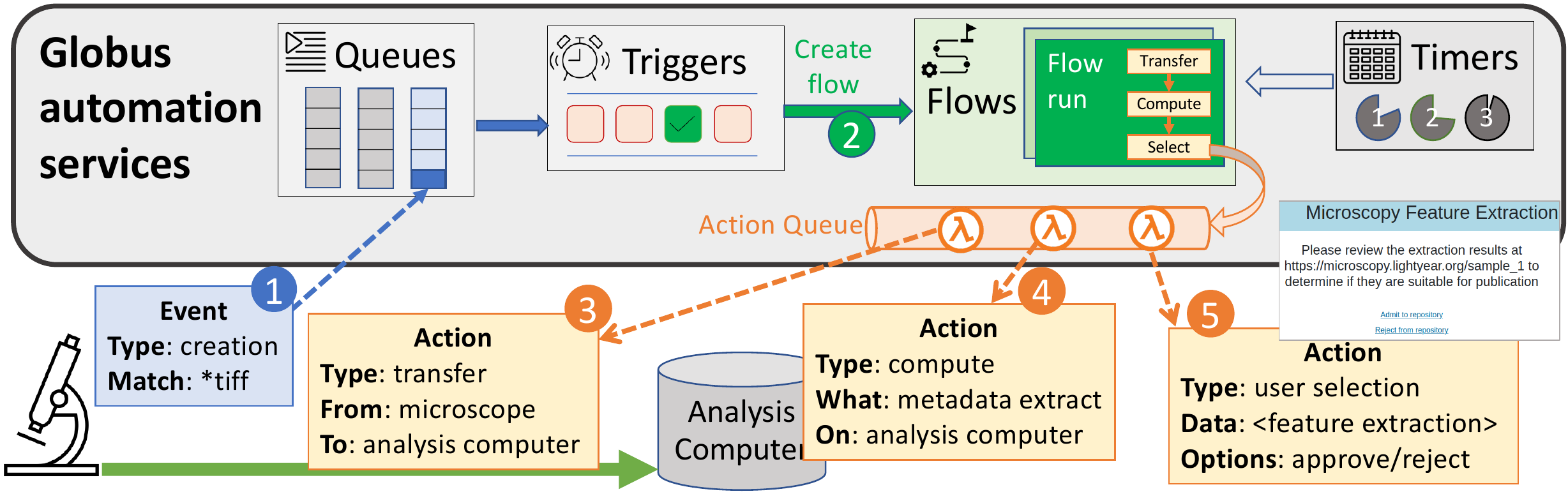}
  \vspace{-2mm}
  \caption{Globus automation services architecture (above) and a simple flow (below). (1) A file creation event at a microscope is sent to Queues. (2) A registered trigger identifies the event and invokes a flow via Flows. The flow invokes three actions, in sequence, each added to the action queue in turn: (3) transfer data from the microscope to an analysis computer, (4) run analysis program to extract features, and (5) prompt the user to approve or reject extracted features. The Timers service (upper right) can also trigger flow invocation.
\label{fig:model}}
\end{figure*}

Based on the model presented above we implement four distinct automation services: 
Flows to create, share, run, monitor, and manage flows;
Queues to reliably store and deliver events;
Triggers to consume events from queues, apply predicates, and invoke flows; 
and Timers to register periodic events. 
We define the action provider API and implement a set of action providers. \autoref{fig:model} shows the main components of the platform.

\subsection{Authentication and Authorization}
\label{sec:auth}

Globus automation services allow users, and agents acting on their behalf, to launch flows that then perform actions on remote services, potentially over extended time periods.
End-to-end authentication and authorization of flows and actions are thus fundamental requirements.
Each request made by a flow to a remote service must provide the credential(s) [in the form of OAuth tokens] that the remote service requires to permit the request.
As the user who launches the flow will not necessarily be available (or have the patience) to provide each such token at that time that it is needed, methods are needed for caching tokens.
These methods must also protect tokens against illicit access and use; circumscribe the purpose(s) for which they can be used; and support their renewal, if needed, in the case of long-lived activities. 

We rely on the OpenID and OAuth~2~\cite{oauth2}-compliant Globus Auth platform~\cite{GlobusAuth} to meet these requirements.
Each Globus automation service, action provider, and flow (collectively, ``service'') is registered with Globus Auth as a resource server and assigned a universally unique identifier (UUID). 
Each such service can, in turn, register OAuth~2 \emph{scopes} representing its operations, each named by a uniform resource locator (URL) that users can employ when granting applications and services consent to invoke the associated operation on their behalf.
A scope URL has the form $<$PREFIX$>$/$<$UUID$>$/$<$OP$>$,
where:
\begin{itemize}
    \item $<$PREFIX$>$ is
\texttt{https://auth.globus.org/scopes/}, indicating that the scope was generated by Globus Auth (as opposed to other OAuth-2 services);
\item
$<$UUID$>$ is the registering service's UUID in Globus Auth (e.g., \texttt{eec9b274-0c81-4334-bdc2-54e90e689b9a} for the Flows service); and
\item 
$<$OP$>$ names an operation on the service (e.g., 
\texttt{publish} and \texttt{manage} in the case of the Flows service; \texttt{run}, \texttt{update}, and \texttt{delete} in the case of a flow).
\end{itemize}
Note that each flow created by the Flows service is itself a service, registered with Globus Auth and having its own UUID, and with its own unique scopes named via the same concatenation of $<$PREFIX$>$, its flow UUID, and operation.

The scope mechanism is central to Globus automation services. It is used to encode fine-grained user consents to authorize clients to invoke specific flows, and to ensure that flows invoke only defined actions.
The Triggers and Timers services use the same model to invoke other services, such as Flows, on behalf of their users.
These fine-grained consents and Globus Auth authorization capabilities together allow Globus automation services to implement a least privilege security model. 

\begin{figure*}
\centering
\includegraphics[width=0.95\textwidth]{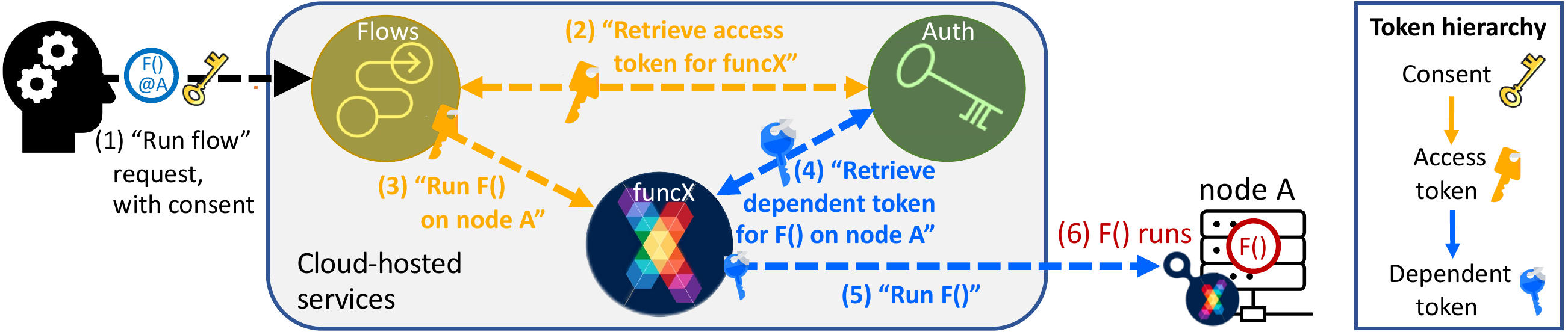}
\caption{Distributed authorization as implemented in Globus automation services. The shaded region in the center encompasses the cloud-hosted Globus services: here, Flows, Auth, and funcX services. (1) A user, having previously authenticated, requests Flows to run a flow comprising the action ``run F() at node A,'' providing the necessary token.
(2) Flows obtains an access token from Auth to make a request to funcX.
(3) Flows requests funcX to run F() on A. 
(3) FuncX obtains a dependent token to run F() on A.
(5) funcX requests the funcX agent associated with node A to run F().
(6) F() runs.
The key on the right illustrates the different tokens used.
}\label{fig:auth}
\end{figure*}

The integration of Globus Auth mechanisms with Globus automation services proceeds as follows.
Each Globus automation service API request must contain an \emph{access token} in the HTTP request ``Authorization'' header. 
Upon receipt of a request, the service uses the standard OAuth introspect operation to communicate with Globus Auth to validate the token and retrieve authentication information, including the caller's identity. 
The information can then be used to authorize operations against policy associated with a flow, as described in \autoref{sec:flow-access-control}. 
As shown in \autoref{fig:auth}, the services may also interact with Globus Auth to retrieve access tokens required to invoke other downstream services (e.g., actions defined in a flow).
Following the standard OAuth-2 protocols, when invoking a service's operation as represented by a scope, a client first requests an access token from Globus Auth. If the client is acting on behalf of a user, the user must grant consent to allow the client to access a specific set of scopes on the service, and any downstream services that the service may need to access.
The downstream services perform the same token introspection to retrieve authentication information, and enforce applicable authorization policy.

\subsection{Action Providers}
\label{sec:APs}

Action providers are the foundation for all work performed by Globus automation services, whether by flows, triggers, or timers. Each implements a common interface for introspecting an action provider's capabilities, requesting execution of an action, and monitoring and managing action progress:

\begin{itemize} 
    \item 
    \code{GET <action\_url>}\code{/}:
    Introspect the action provider for descriptive and administrative information, the required Globus Auth scope for invocation, and schema for the action's input. This operation may be permitted without any authentication allowing, for example, the scope to be discovered without otherwise authenticating or possessing access tokens.
    
    \item 
    \code{POST <action\_url>}\code{/run}:
    Invoke an action described by a supplied input document that matches the schema returned from introspecting the action. The request contains a client-generated \code{request\_id} which is used by the action provider to de-duplicate repeated requests. This operation returns a document containing an identifier for the new run (the \code{action\_id} used in subsequent requests), the state of the action (\code{ACTIVE}, indicating still running; \code{SUCCEEDED}; or \code{FAILED}), and action-specific details on the state or result of the run.
    
    \item 
    \code{GET <action\_url>/<action\_id>}\code{/status}: 
    Retrieve, for the run with the provided \code{action\_id} value, a document with the same form as that returned from the \code{run} operation.
    
    \item 
    \code{POST <action\_url>/<action\_id>}\code{/cancel}: 
    Request cancellation of a run in the \code{ACTIVE} state. Cancellation is considered advisory only: it may stop the run immediately, cause the run to end sooner than normal, or have no effect. In all cases, a document is returned with the same form as that returned from the \code{run} operation structure, to report the (potentially updated) status of the run.
    
    \item 
    \code{POST <action\_url>/<action\_id>}\code{/release}: 
    If run has completed (i.e., state is \code{SUCCEEDED} or \code{FAILED}), remove its context from the action provider. 
    (Action providers typically otherwise retain state for 30 days.)  
    The same action status information is returned as for the other operations, but upon completion, any subsequent references to the \code{action\_id} will be unrecognized by the action provider.
\end{itemize}

The Flows service implements this API for each flow at the URL \texttt{https://flows.globus.org/flows/<flow\_uuid>}.\\ Thus, anywhere that a Globus automation service may invoke an action provider, as an action within a flow or via the Triggers or Timers services, it can also invoke a flow.

To enable users to create and operate new action providers, either for their own use 
or to share with others, we provide a Python library that defines classes for the common Globus Auth and action provider data types and operations \cite{actionprovidertoolkit}. This library further provides helpers for setting up the required REST API entry points for common Python web development environments. We use this development kit in developing the action providers enumerated in \autoref{sec:actionsusage}.

\subsection{The Flows Service}

Our Flows service implementation has two main components:
\begin{inparaenum}[(1)]
\item A horizontally scalable front-end service that implements the REST API for publication, invocation, monitoring, and other lifecycle operations of a flow; and
\item a back-end polling process that initiates and monitors the progress of actions by using the Action Provider interface described above.
\end{inparaenum}
In addition, we make extensive use of AWS services to 
provide the scale, reliability, and availability required by a large user community with critical use cases. 

\begin{figure*}[ht]
  \vspace{-0in}
  \centering
  \vspace{-2mm}
    \includegraphics[width=\textwidth, clip]{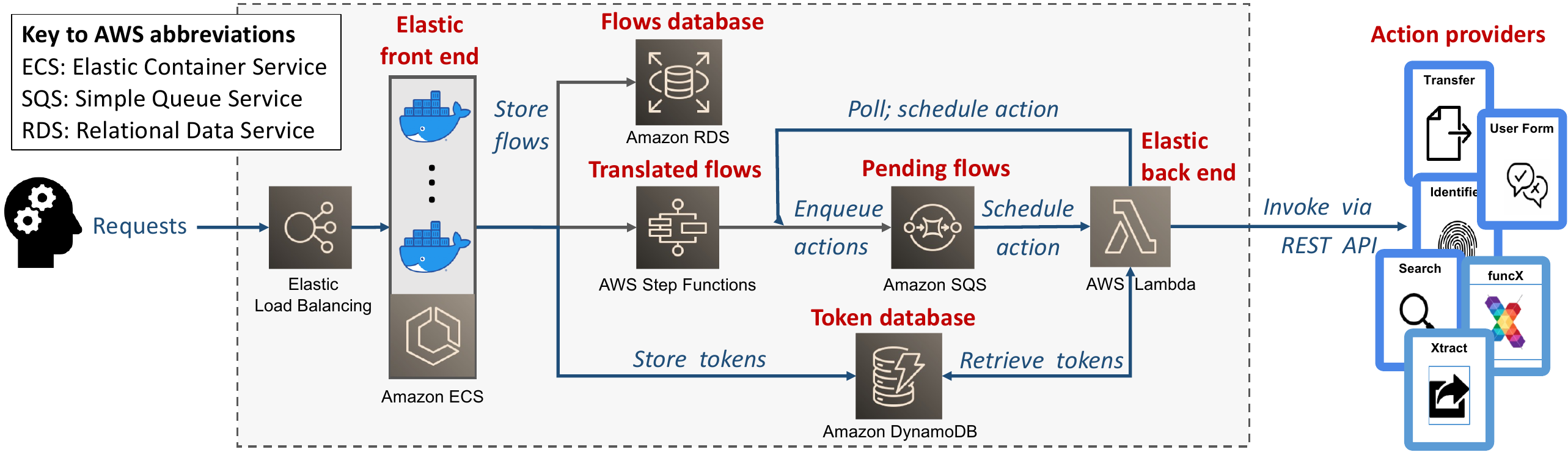}
  \caption{The Flows service front end is deployed in Docker containers, managed by AWS ECS. A variety of other AWS services, shown as grey-scale icons with the service name underneath, are employed for persistence, flow execution, and reliable scaling of interactions with remote Action Providers. The text above each AWS service indicates its function.}
\label{fig:flowsblockdiagram}
\end{figure*}

The structure of the service, including its front end and back end, and its use of AWS services, is shown in \autoref{fig:flowsblockdiagram}. The front end implements all client facing REST APIs for life-cycle management of flows and runs. We deploy the front end in Docker containers using the AWS Elastic Container Service (ECS) for automatic scale-up and scale-down based on demand. The front end stores state in a replicated Postgres AWS Relational Data Service (RDS) database cluster, and shares authentication state with the back end via an AWS DynamoDB table. The other Globus automation services described here, Queues, Triggers and Timers, make similar use of AWS for hosting a scalable API front end, database services for persistence, and in the case of Triggers and Timers additional back-end processes for monitoring actions they initiate.

Flow runs are passed to another AWS service, AWS Step Functions (ASF), which in turn invokes the Flows back end via an AWS Simple Queuing Service (SQS) message paired with an AWS Lambda function that implements the Flows back end. By performing all execution and monitoring of flow runs and their action steps outside the front end, we ensure that flows will continue to make progress even if the front end is down and that this execution environment will scale should many runs be in flight simultaneously.

\subsubsection{Flow deployment}
\label{sec:flow-deployment}

The most complex elements of the Flows service are those concerned with the deployment and execution of flow runs. 
Deployment of a flow requires interaction with two platform services, Globus Auth and AWS Step Functions (ASF).

As described in \autoref{sec:auth}, the Flows service registers each newly deployed flow with \textbf{Globus Auth} as a separate
service, and further registers new scopes for the operations used to run or manage the flow. 
Prior to registering each scope, the
service examines the flow definition to identify all action
providers that it may use, and makes each of the action provider's scopes dependent on the scopes for this flow. Thus, when a user runs the
flow, they will be informed of the action providers they are allowing
to be invoked on their behalf and with their identity during the flow's run.

The interactions with \textbf{ASF} involve deploying a new ``state machine'' (the ASF equivalent of a flow) for each flow. This state machine is a transformed version of
the input flow received by the service. The transformation involves two components:
\begin{inparaenum}[(1)]
\item generating ASF ``Task'' states for each Action in the flow and;
\item altering references to the flow's run-time state to protect service-specific data stored at run-time.
\end{inparaenum}

ASF Task states are used to perform activities, and they may be used to invoke a wide variety of AWS resource types. The Flows service creates Task states which pass the information related to an Action step onto an AWS SQS queue for which a Lambda function is configured as a receiver. The queued messages include the URL for the desired Action Provider, the name of the scope used to invoke the action, parameters to the action invocation, timeout and error-handling options.

The Flows service stores additional information for its own use, such as identifiers for the flow, run, and invoking user in the state of the ASF state machine execution context. To protect this internal information from the user, all initial user input and references to run-time state are stored in the context under a key \code{\$.UserState}. Thus, all references within all states of the flow definition are updated with this prefix so they will be read and written to this sub-key of the flow's run-time state.

\subsubsection{Processing Flow runs}
\label{sec:flow-runtime}

When a user invokes a flow, the user's identity is confirmed via Globus Auth, and authorization is granted if that identity satisfies the ``Runnable By'' policy configured for the flow. Tokens identifying the invoking user for all dependent scopes established when the flow was deployed are retrieved from Globus Auth. 
This process is repeated for any additional roles used in the flow. Once retrieved, tokens are placed into the DynamoDB database for use when interacting with action providers on the user's behalf during the run. ASF is then invoked to start the run with the user's input, as well as additional information described above are placed into the into state of the run Context. From this point on, the flow run proceeds without interaction with the Flows service front end. Thus, should the front end be off-line, all runs can nevertheless proceed correctly.

The transformed flow that is deployed to ASF causes each action step in the flow run to invoke the back-end Lambda function used to interact with the appropriate Action Provider.
The Lambda function uses the action provider URL and access tokens from the token database to invoke the action provider. It passes the action's message body
as arguments to the action provider and receives an \code{action\_id} in response. Subsequent invocations of the
Lambda function use the \code{action\_id} to retrieve the status of the invocation, determine its completion or failure (including enforcing a timeout), call action provider
\code{release} to free resources, and return the invocation's result to ASF. When a poll of the action returns an incomplete state, the polling period is updated and the message is returned to the queue
with an updated polling period.
Specifically, an initial period is set in the invocation message body.
That period
is then doubled each time the status is checked up to a maximum of 600~s. Thus single action queue is used to process both initial and polling requests for an action invocation, with the queue itself providing the delay between polls.

\subsection{The Queues Service}
The Queues service supports the creation and use of \textit{queues}, which Globus automation services use for reliable and secure delivery of messages 
from \textit{senders} to \textit{receivers}---specifically, from event generators to triggers.
It allows for asynchronous communication: events can be added to a queue, and will be stored, even if no active receiver is currently associated with the queue, or if the receiver is temporarily incapable of receiving and processing messages at a rate matching that of the queue's sender(s). It also ensures in-order message delivery.

The Queues service is implemented as a thin layer over Amazon SQS, with each user-created queue realized as an SQS queue. The Queues service augments SQS by integrating it with the Globus Auth identity and access model. 
The Queues REST API defines methods for creating, modifying, and deleting queues---and, once a queue is created, for adding, receiving, and acknowledging receipt of messages to/from that queue. 
Three roles associated with each queue control who can modify policies or delete the queue (Administrator role), send a message onto the queue (Sender), and retrieve a message from the queue (Receiver).

The Queues service implementation uses message receipts to provide \textit{at-least-once} message delivery semantics: 
Each message received from a queue includes a unique message identifier, and only after that identifier is returned to the queue in a subsequent acknowledgement API call can the message be removed from queue storage.
If no acknowledgement is received after a certain period, the message may be re-delivered. 
In addition, message identifiers are used to ensure \textit{exactly once} invocation semantics: Each time that the trigger service invokes an action, it uses the queues service message identity as the request identifier, and as described in \autoref{sec:actionsusage} action providers will discard requests with duplicate request ids.

\subsection{The Triggers Service}
The Triggers service enables configuration and execution of triggers, which are used for event-based invocation of a flow or action.
The creation of a trigger is a two-step process. First, the user interacts with the Triggers service to \textit{configure} a new trigger. In doing so, the user provides the identifier for a previously created queue, the URL for the action provider (which may be a flow) to be invoked by the trigger when an event occurs, a predicate used to identify events that should cause the action to be invoked, and a transformation to be applied to each triggering event's properties to create the input for the resulting action.

Second, the user then requests that the Triggers service \emph{enable} the newly created trigger. 
In so doing, 
the user provides an access token with dependent scopes for the Queues receive message and for running the action, so as to grant the Triggers service authority to read from the queue and to invoke the action.
A user may also \emph{disable} a trigger, which places it in an idle state in which no events from the queue will be processed.

While a trigger is enabled, the Triggers service periodically polls the queue.
Polling is performed in the service's back-end 
by a pool of workers that select enabled triggers from a priority queue that encodes the time until the next poll should occur. A trigger is placed back onto the queue after polling, increasing the polling interval when no messages are available and decreasing the interval when messages are received.

For each message received, the trigger's predicate is evaluated to determine if a match occurs. The predicate is a Boolean expression written in a Python-like syntax that may evaluate any properties of the incoming message. For example, if the message represents a file creation event, the predicate may check that the filename ends with a particular suffix, such as ``.tiff''. Messages that do not satisfy the predicate are discarded.

Those messages that satisfy the predicate will cause the trigger's configured action to be invoked. The input to the action is formed from the properties of the incoming message. The various parameters are specified using the same Python-like syntax as the predicate and can evaluate properties of the incoming message. For example, if a filesystem update event message contains a list of new files called simply \texttt{files}, but an action needs an input parameter \texttt{number\_of\_files} the transformation could be written as \texttt{number\_of\_files = len(files)}.
Once the input is formed, the action is invoked using the access token acquired when the trigger was enabled. The run's identifier (\texttt{action\_id}) is added to a queue so that the same polling process can be performed to monitor the progress of each run. When the run completes, its results are cached in the trigger's configuration so that recent results and statistics related to the trigger's usage may be retrieved by the user.

\subsection{The Timers Service}
The Timers service has a similar purpose and a similar internal structure to the Triggers service. Whereas Triggers invokes actions in response to events, Timers invokes actions at regular time intervals. 
The configuration of a timer includes: 1) the identifier for an action (which may be a flow) to be invoked; 2) start time for action invocation; 3) a time interval in seconds; 4) either a count of the number of times to invoke the action or an end time; and 5) input arguments for the action. The Globus Auth scope for creating a timer is dependent on the action or flow scopes; thus, the Timers service retrieves an access token when the timer is configured, and uses it to invoke the action.

Internally, the Timers service is implemented similarly to the Triggers service: when a timer is established, its start time is inserted into a priority queue sorted by timestamps for next execution time. A single back-end dispatcher process wakes periodically and pops any element(s) from the queue whose next time is less than the current time. For each timer thus identified, it posts an invocation request onto a separate work queue, computes the timer's next execution time using the defined interval, and places it back onto the work queue as long as it will not have expired based on the count or stop time parameter. A set of worker processes listen on the work queue, and for each timer received, use the action parameters and the access token required to invoke the action. As queues are maintained with persistent storage, timers are not lost if the Timers service is down: once the service restarts, it will recover any missed timers and schedule the required actions.
Timers are currently only available in the Globus platform and web application to perform periodic data transfers; however, the architecture and implementation support the invocation of any action.

\section{Evaluation}
\label{sec:evaluation}
    
We first investigate the performance of Flows and the latency and overhead involved in executing individual flows. Then, we consider the performance of the action providers. Finally we review use and adoption of Globus automation services in production settings.

\subsection{Flow Throughput and Latency}

To examine the throughput of the Flows service and its ability to serve many users concurrently, we performed load tests in which an isolated instance of the service, deployed on a single ECS container with a CPU value of 2048 (equivalent to two vCPUs), 4~GB of memory, and eight worker threads, served requests from varying numbers of clients.
The clients were deployed on a login node of Argonne's Theta computer with an Intel Haswell E5-2698 v3 CPU with 256~GB of DDR4 memory.
We then performed experiments in which each of $N$ concurrent clients, for $N$ = 1, 2, \ldots, 128, repeatedly invoked a simple flow comprising a single Pass state (essentially a no-op) and waited for the response. We measured both the time from invocation to response for each request (latency), and the average number of requests processed per second (throughput).

\begin{figure}[ht]
  \vspace{-0in}
  \centering
  \includegraphics[width=\columnwidth,trim=5mm 6mm 5mm 5mm, clip]{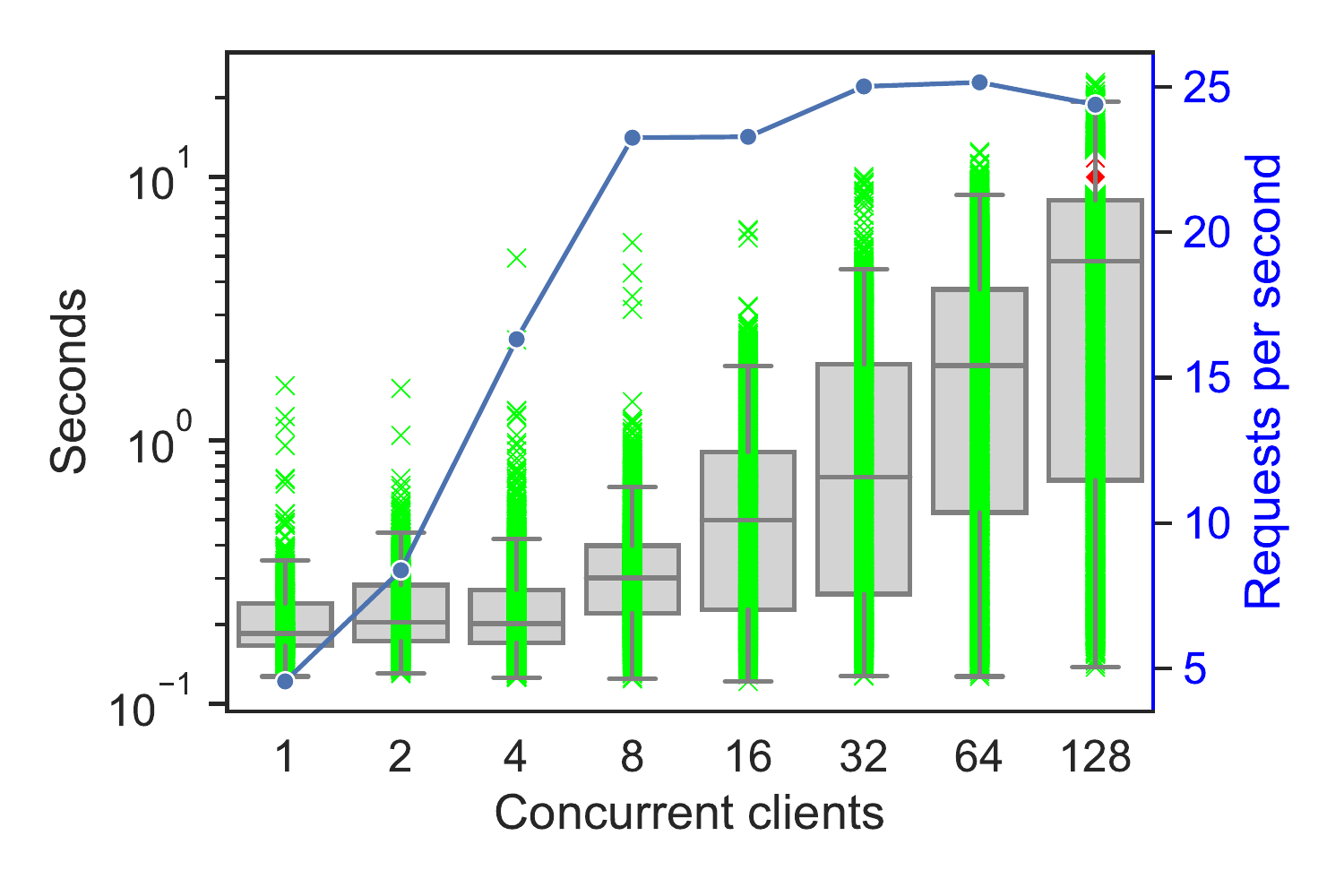}
  
  \vspace{-2mm}
  
  \caption{Flows service performance when processing requests to run a simple flow.
  X-axis: Number of concurrent clients making requests.
  Left y-axis: Request response time, plotted both for individual requests (green cross for success, red for failure) and as box plots for lower and upper quartiles, with whiskers to 1.5$\times$ the interquartile range.
  Right y-axis: Requests per second.
  }
\label{fig:throughput}
  \vspace{0in}
\end{figure}

We see in
\autoref{fig:throughput} that the Flows service in the measured configuration can serve roughly \num{25} flow invocations per second when under load, with failures appearing with more than 64 concurrent requests. The number of requests per second plateaus once eight concurrent clients are used, as the many clients begin to saturate the eight available worker threads. 
Failures occur under high load because each of the service's worker threads is busy communicating with the ASF service, meaning the load balancer is unable to pass the request to the service. Such failures can be avoided by dynamically scaling the number of instances deployed by Flows. 
We note that the production Flows deployment employs a minimum of four containers and can scale further horizontally, based on load.

\begin{figure}[ht]
  \centering
  \includegraphics[width=\columnwidth,trim=0.07in 0.08in 0.07in 0.08in,clip]{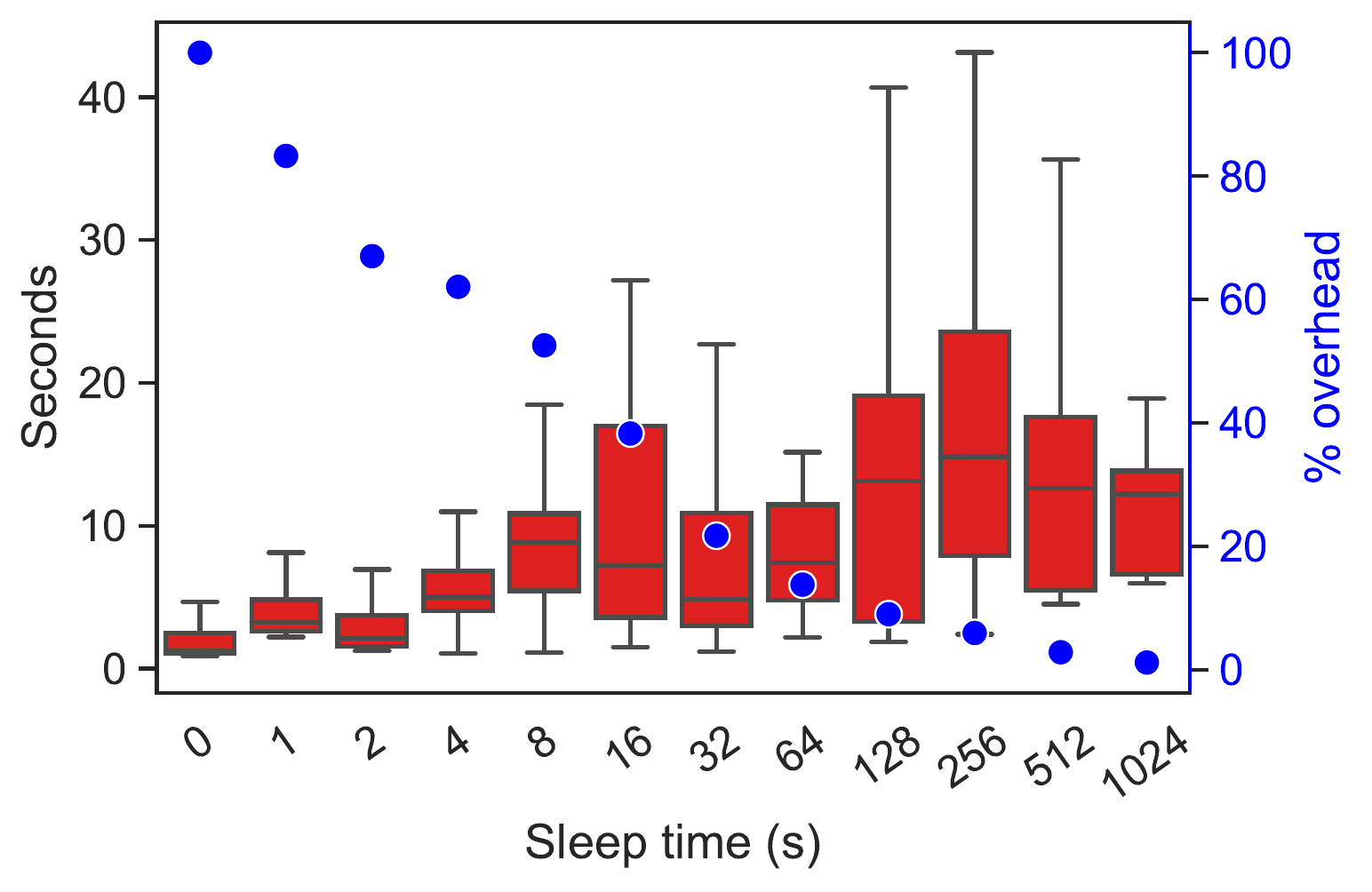}
  
  \vspace{-2mm}
  
  \caption{Overhead incurred by a single-step flow with a sleep action of a specified duration (x-axis). The left y-axis and red box plots show the overhead in seconds, with upper and lower quartiles. The right y-axis and blue markers show mean \% overhead.}
\label{fig:flowoverhead}
\end{figure}

In a second set of experiments, 
we ran a flow consisting of a single action that sleeps for a specified period of time and measure the overheads associated with flow execution, which we define as flow completion time minus the action sleep time.
\autoref{fig:flowoverhead} shows that no-op flows (sleep time of 0~s) incur, on average, \num{2.88}~s overhead. 
This cost is due primarily to the exponential backoff policy used by the Flows service when polling for task completion: 
Each task is first polled after 2~s and, each time that the task is found still to be active, the polling interval is doubled, up to a maximum of \num{600}~s.
The remainder of the overhead is due to the polling request being queued for processing by a Lambda function and the cost of communicating with the remote action provider. 
The figure also shows that flow overheads as a fraction of total flow time decline as flow runtimes increase, to an average of 1.2\% for \num{1024}~s flows.

\subsection{Action Providers}
\begin{figure}[ht]
  \centering
  \includegraphics[width=\columnwidth,trim=3mm 3.5mm 2.5mm 2mm,clip]{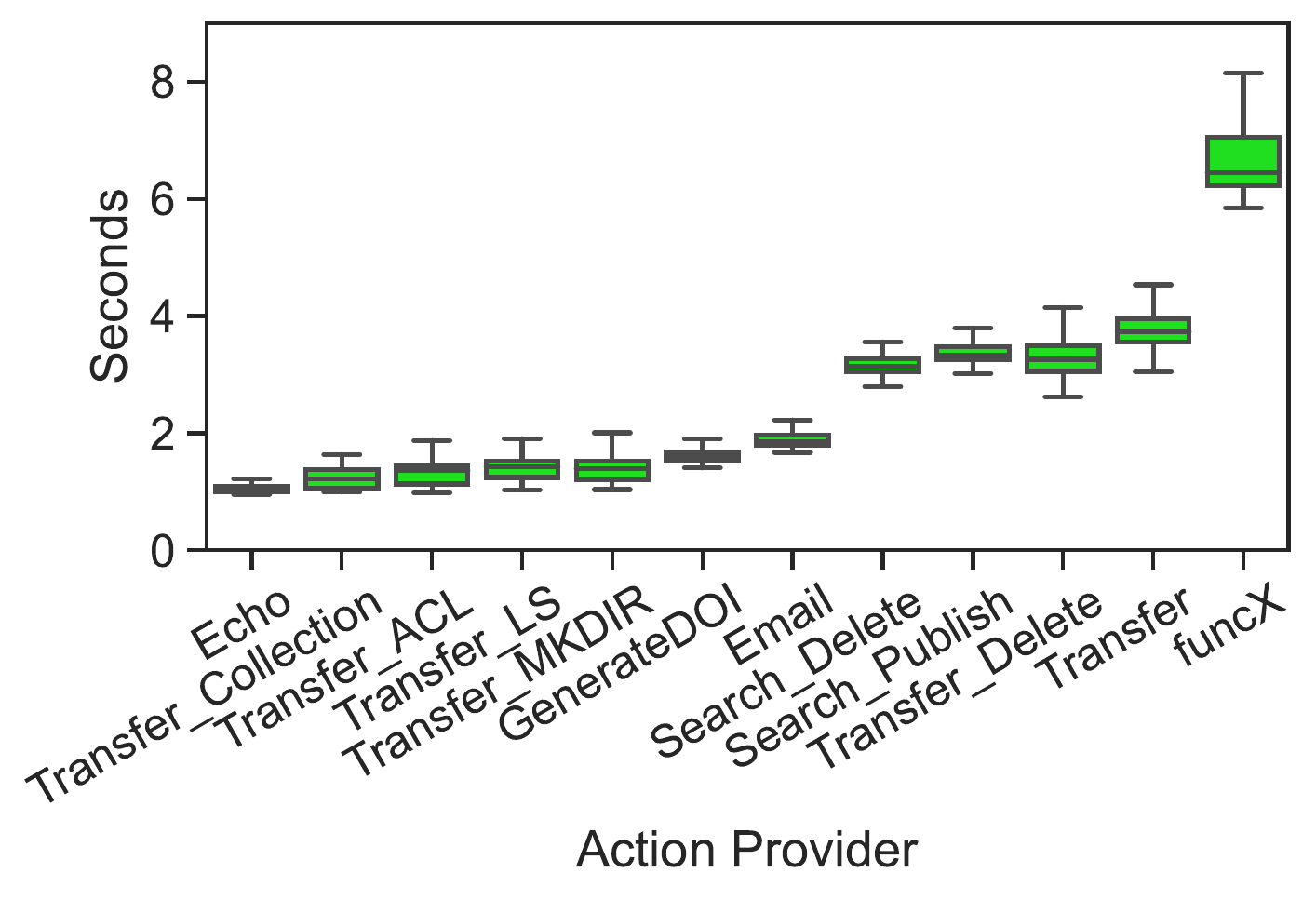}
  
  \vspace{-2mm}
  
  \caption{Round-trip latencies observed for various action providers, each executed at least 100 times. For Transfer and Search, we separate out results for different action options.}
\label{fig:aplatency}
\end{figure}

Understanding action provider performance is crucial to developing efficient
flows and choosing timeout values. Thus, we performed experiments to measure round trip latencies for various actions.
In each case, the requested action involves a simple task: e.g., transfer a four-byte file, run a no-op function, and index a trivial record into a search catalog. Thus, the measured costs, shown in \autoref{fig:aplatency}, are largely overhead associated with negotiating access to the corresponding service. We do not evaluate here costs that scale with, for example, the size of the data being transferred or published. 

We see in \autoref{fig:aplatency} that simple tasks, such as Echo, are completed relatively rapidly, albeit with a $\sim$1~s floor on response time.
More demanding actions, such as funcX and data transfer, take longer. 
Analysis suggests that these higher costs are due to administrative overheads. 
Authentication accounts for around 200--400~ms of a typical request. 
In the case of funcX, a majority of the time is spent instantiating a secure client to interact with the funcX service---a cost that is amortized if multiple functions are bundled in one request.

The relatively high action execution times seen in these experiments preclude certain applications of Globus automation services.
However, we have been pleasantly surprised by how many research automation applications can function effectively under these parameters. 
The reduction of various overheads, for example by caching credentials and proxy clients, will be a focus of future work.

\subsection{Production Flows}

\begin{figure}[ht]
  \vspace{-0in}
  \centering
  \includegraphics[width=\columnwidth,trim=6mm 6mm 6mm 6mm,clip]{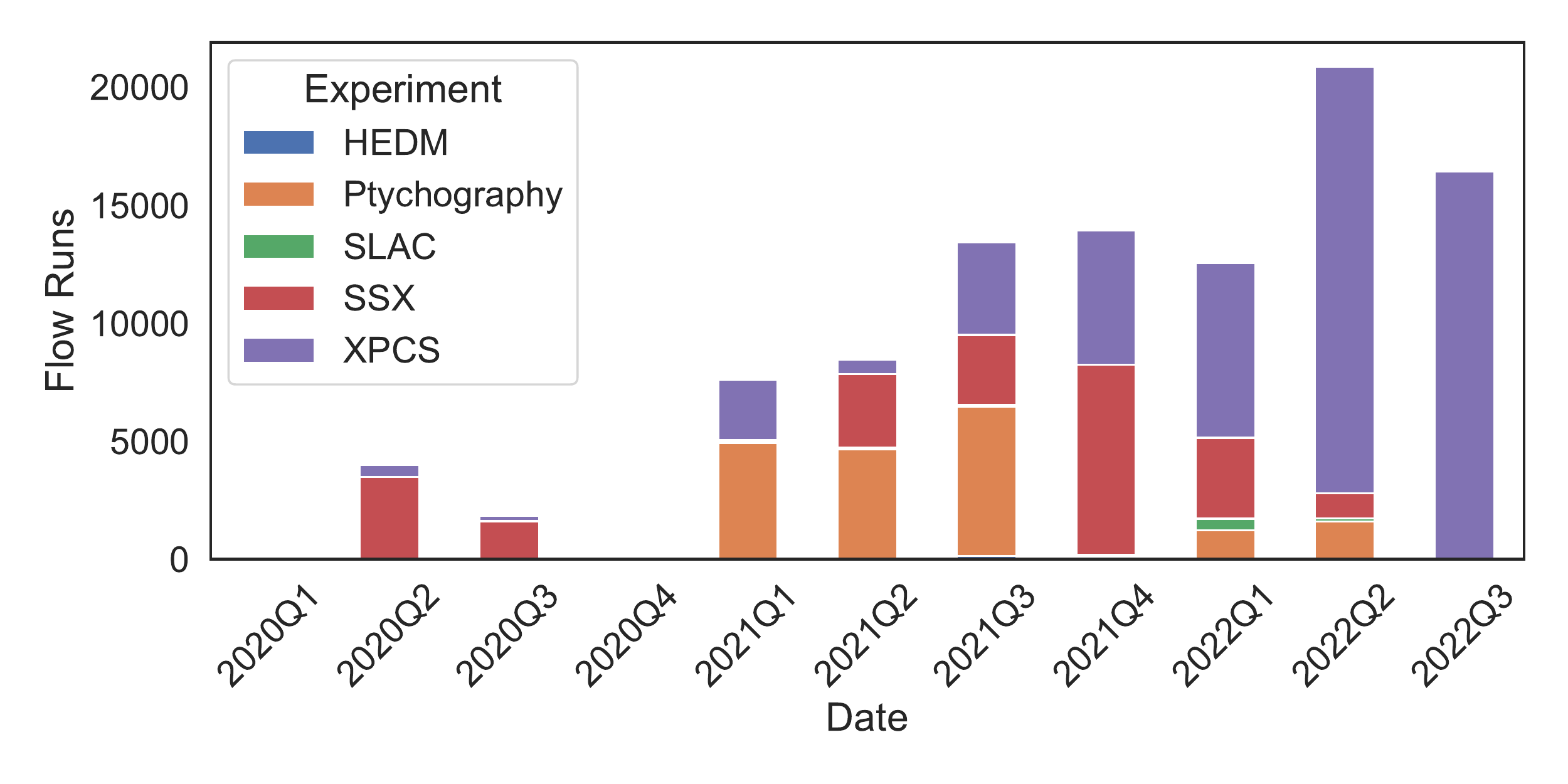}
  
  \vspace{-2mm}
  
  \caption{Number of flow invocations over time from five different APS experiments. The numbers vary due to facility and experimental schedules. 
  The decline in the latest quarter is due to APS preparing for an extended shutdown for upgrade.
  }
\label{fig:apsusage}
\end{figure}

Globus automation services are increasingly being used to run production workloads
and are indeed becoming an integral part of many research data lifecycles.
One facility that leverages the platform is Argonne's Advanced Photon Source (APS), a synchrotron light source facility that houses 68 beamlines in 32 sectors used by more than 5000 scientists a year. 
Since prototype Globus automation services were first made available in 2020Q1, 
adoption has grown from a few experiments to thousands of
flows that are used routinely to analyze and catalog experimental data. 
This adoption is shown in
\autoref{fig:apsusage}, which summarizes usage of the services across five APS beamlines.

We review a set of 415 production flow runs performed between December 9th and 15th, 2021, in support of the experimental science use case of \autoref{sec:usecases}.
All runs involved the same flow, which comprise a total of six steps (see \autoref{tab:xpcs}) used to retrieve and analyze an experimental dataset, generate images, and publish results to a search catalog.
Each individual run was triggered by the creation of a new dataset at the experimental facility.

\begin{table}[b]
    \centering
    \caption{Times, in seconds, to process the steps of 415 flow runs used to analyze and publish datasets from an APS experiment.}
    \label{tab:xpcs}
\begin{tabular}{lrrrr}
\toprule
Action & Min & Max & Mean & Std \\
\midrule
Transfer & 4.11 & 522.66 & 47.61 & 95.95 \\
Pre-publish & 3.50 & 44.19 & 7.01 & 5.71 \\
Analyze & 7.54 & 2881.93 & 326.17 & 487.01 \\
Visualize & 20.03 & 549.50 & 116.71 & 98.30 \\
Extract & 6.65 & 52.51 & 10.94 & 5.53 \\
Publish & 3.64 & 34.54 & 7.44 & 4.88\\
\bottomrule
\end{tabular}
\end{table}

In total, the 415 runs processed over 500~GB of X-ray 
imagery and consumed over 1500 supercomputer node hours. 
The dataset generation rate at the instrument, and therefore the rate at which 
the flow was invoked, ranged from 0.1 to 0.0001 Hz,
depending on the collection technique in use and beamline operational procedures. 
\autoref{tab:xpcs} characterizes the times taken by the six states over the 415 runs.
The large variations are due to:
1) changes in data collection technique over the course of the experiment, which resulted in data sizes that varied by two orders of magnitude, and thus varied transfer and analysis times; and
2) resource contention at times of peak collection rate, which led to both transfer and execution tasks being queued by either the transfer service or the HPC scheduler.
Nevertheless, every dataset collected during this period was successfully processed and published to the search catalog.

\section{Experiences and Lessons learned}
\label{sec:cases} 

From when Globus automation services were first released in beta in 
2021, to August 2022,
84 unique users have defined 4737 flows
and 167 users ran 
\num{247643} flows,
of which \num{225162} either ran to completion successfully or are active at time of writing,
\num{20189} failed (typically due to timeouts, as noted earlier),
and \num{1971} were cancelled.
A further \num{321} flows are ``inactive'', meaning that they
have stalled for various reasons (e.g., expiry of credentials required to transfer data).
There are now 14 separate action providers. 
The Timers service has been used by 813 users to create 3642 timers that have cumulatively
fired \num{1777271} times; 338 timers from 188 users were active at the time of writing. 
The Triggers and Queues services are prototypes and are not yet widely accessible.  

As Globus automation services have evolved from prototype to production, 
we have worked with various groups to 
define and deploy flows. Early adopters were
primarily from the four main use cases
outlined in \autoref{sec:motivation}. 
We have seen significant adoption, in particular, in instrument science, where most flows authored to date follow a pattern in which 
preliminary data processing is performed near the instrument, data are then moved to a compute cluster, further analysis is performed on that cluster, and results are returned to scientists, 
either directly or via a web-based catalog. 
We note the following characteristics
of these early uses:

\vspace{-1mm}

\begin{itemize}
    \item 
    \textit{Flow diversity}\/: Even in situations in which the high-level process appears similar, implemented flows must be adapted to specific use cases. For example, 
    the instrument use cases have similar processes, yet each flow has different 
    actions and configurations.
    
    \item \textit{User diversity}\/: 
    Globus automation service users range from 
    software developers to scientists with limited programming
    experience. The various Globus automation services interfaces satisfy the requirement for different
    ways of working with the platform, but further
    abstractions are needed to reduce barriers for non-expert users. As such, we are actively working on both a Python toolkit~\cite{vescovi2022linking} and graphical interface to compose flows.
    
    \item 
    \textit{Throughput over latency}\/: 
    Our current applications have not needed sub-second responses; rather, users are concerned with high throughput, and thus with being able to process many flows in parallel, as well as with reliability.
    
    \item 
    \textit{Authorization bottlenecks}\/: The authorization steps required for flows to access resources in different administrative domains can be a source of complexity. Our approach provides a structured way of managing authorizations, but does not overcome the need for periodic refresh of consents and credentials---a task that can become tiresome, particularly when different user identities must be employed on different resources. In some settings, group-based access controls can be a solution. 
    
    \item \textit{Provenance}\/: 
    Researchers often want to review flow executions, understand under what conditions
    a flow was run, explore why a flow failed, and review performance and other metadata regarding individual actions.
    
    \item 
    \textit{Flows are logically grouped}\/: Many use cases execute multiple flows as part of a single unit of work, such as a particular experiment or data publication task. Users want to think about and manage such a collection of flows as a unit, with the ability to drill down into the details of individual flows where necessary. To this end we have implemented \emph{tags} to filter collections of runs and simplify discovery. Providing additional grouping, discovery and filtering capabilities is an area for future work.
    
    \item 
    \textit{Action provider API implementation is a significant undertaking}\/: Our approach of implementing action providers as standalone services and as wrappers around existing services using Lambda functions has required custom optimizations in terms of token caching, horizontal scaling, and resource configurations. Higher-level abstractions are needed to make it easier and faster to develop new action providers.
\end{itemize}

The Globus automation services approach to automation is \emph{not} intended for:
1) \emph{computational workflows involving many tasks},
for which specialized workflow tools (\autoref{sec:related}) exist;
2) \emph{high-volume flows}, 
such as responding to every change in a file system
(Globus automation services are designed to process millions, not billions, of flows per day,
and rate limiting is used to prevent denial of service attacks on the services); and
3) \emph{workflows where high performance or millisecond-scale latency matters}\/: 
for example, for complex event processing on data streams from real-time monitors.

\section{Related Work}
\label{sec:related}

Hundreds of workflow systems have been created to orchestrate sequences of steps~\cite{workflowlist}, with goals of automating computational campaigns, making efficient use of parallel or distributed systems, and representing complex processes~\cite{ludascher2006scientific,goecks2010galaxy,deelman2015pegasus,albrecht2012makeflow,babuji19parsl, silva21workflows}. 
Several authors have attempted to organize and categorize 
the many workflow systems~\cite{liew2016scientific, krauter02gridmanagement, deelman2009workflows}.
We considered various of these existing systems before 
deciding to build upon ASF. We briefly
review different workflow approaches. 

\textbf{Workflow task models}:
Workflow systems are used broadly to coordinate different types of activity, such as local programs, jobs submitted to parallel computers or clouds, calls to web services, and human activities.
Task-based systems, such as Parsl~\cite{babuji19parsl}, Pegasus~\cite{deelman2015pegasus}, and Swift~\cite{swift} execute computational tasks, either by invoking program functions or by making calls to locally executable programs and scripts. 
Service-based systems, such as Taverna~\cite{hull2006taverna}, Netflix Conductor~\cite{conductor}, and ASF, are designed to invoke web services, for example via Web service protocols~\cite{curbera2002unraveling}: what is sometimes referred to as microservice orchestration~\cite{alshuqayran2016systematic}. 
Some systems, such as HyWare~\cite{candela2021workflow}, track automated and human-based tasks. 

\textbf{Workflow representations}:
Workflows may be represented either declaratively or imperatively.
In declarative systems, a structured notation (e.g., a simple textual notation in DAGman \cite{dagman}, a JSON- or YAML-based notation in the Common Workflow Language~\cite{cwl}, and an XML-based notation in Pegasus) is used to specify the actions to be performed and their relationships; an interpreter or compiler then translates this specification into runtime operations.
In imperative systems, a workflow is implemented by an executable program coded with extensions to an existing language (e.g., Parsl extends Python) or an independent domain-specific language (e.g., Swift). 
We adopt a declarative representation in this work (specifically, an extended version of ASL) due to the convenience of a textual representation that can be generated variously by libraries, graphical user interfaces, or compilers.

\textbf{Workflow deployment models}:
Most workflow systems are designed to be run by a single user in order to execute that user's workflows.
Some systems (e.g., Galaxy~\cite{goecks2010galaxy}, Taverna)
can be deployed in a multi-user model via which
groups of users may define, share, and manage workflows . However, in most cases these workflow systems are deployed on a single computer by those who use them. 
Hosted workflow systems, such as those offered by cloud providers and the public Galaxy instance, allow users to define and execute workflows without installing and managing workflow systems locally. 
However, these systems are typically bound to the cloud platform or cluster on which they are deployed. 
We build upon this model to provide a hosted service via which users can outsource the execution of workflows to a trusted and reliable third party.
    
\textbf{Business process automation}:
Business process automation systems seek to represent enterprise information processes in executable forms, for example via the Business Process Execution Language (BPEL)~\cite{bpel}, which allows for the linking of web services 
and human processes.
These systems were not designed for research process automation and there is only limited experience with their use in science~\cite{emmerich2005grid,tan2010comparison, tan07bpel4job}.

\textbf{Cloud services}:
Cloud providers are delivering many innovations 
in automation, for example, for development operations (DevOps), combining different cloud services, and data-oriented workflows. 
These services often focus on higher-level automation goals by chaining existing cloud services; increasingly, they aim to lower barriers to use. 
Thus, for example, AWS provides both a full-featured Simple Workflow Service (SWF)~\cite{simpleworkflowservice} and a simpler Step Functions (SFN) service~\cite{amazonstepfunctions}. 
Software development services, such as GitHub Actions~\cite{githubactions} and Amazon's CodePipeline~\cite{codepipeline}, provide automation tools forx
continuous integration and continuous deployment processes.
These tools enable users to combine actions into pipelines that perform DevOps tasks in response to code events.

\textbf{Event-based models}:
The use of queues in Globus automation services to link event producers and consumers reprises the pub/sub model often used in distributed systems~\cite{eugster2003many} and sometimes in scientific workflows~\cite{alqaoud2009publish,kamburugamuve2015framework,renart2017online}.
EPICS~\cite{epics} and ROS~\cite{quigley2009ros} use this model to control experimental and robotic systems, respectively. 
The integrated Rule-Oriented Data System (iRODS)~\cite{xu2017irods} enables specification of data-related processes.
Trigger-action programming~\cite{ur2014practical} seeks to create
user-friendly interfaces for creating automations. If-This-Then-That (IFTTT) \cite{ur2016trigger} allows users to select events and actions via a graphical interface, for example to turn on lights at specific times or control a thermostat based on proximity. These concepts have also been applied to scientific data~\cite{ryanref}.

\textbf{Remote computing interfaces}:
Many customized solutions have been developing for linking scientific instruments with HPC, for example in biomedicine~\cite{goscinski2014multi}, environmental science \cite{plale2006casa,elias2017s}, and disaster response~\cite{beckman2007spruce,altintas2020using}; using HPC to analyze large data~\cite{boccali2021dynamic}; and providing on-demand access to HPC~\cite{wilkins2008teragrid,blaschke2021real}.
Such applications have motivated the development of specialized interfaces for remote job submission~\cite{cholia2010newt,stubbs2021tapis}
and for managing workloads across systems~\cite{thain2005distributed,salim2019balsam,nickolay2021towards,antypas21enabling}.
The LBNL superfacility project has studied requirements for linking instruments with HPC~\cite{bard2022lbnl} and proposed an OAuth-based 
API~\cite{bard21superfacility} that is similar to our action provider interface. 
DataFed~\cite{stansberry2019datafed} federates various scientific data stores.

Researchers investigating methods for autonomous scientific discovery~\cite{sparkes2010towards,roch2018chemos,steiner2019organic,burger2020mobile,stach2021autonomous,noack2021gaussian} have developed innovative approaches describing discovery protocols, but have not yet addressed the systems issues encompassed by Globus automation services.

\section{Conclusions}
\label{sec:conclusion}

The work reported here has been motivated by ongoing investigations of how best to automate currently manual research processes. 
We developed Globus automation services to address these unique requirements, enabling users to define, publish, share, and invoke flows composed of various external actions. 
We developed a declarative flow representation building upon ASL; an asynchronous action provider API to enable integration of various actions; and a robust and scalable set of services to manage the secure invocation and execution of flows. 
We integrate a flexible authorization model via which
flows, Globus automation services, and actions are registered as independent OAuth~2 resources, such that users may delegate authorization to these components to manage the secure invocation of flows that span a wide range of temporal and spatial extents.
Further, 
having identified reliability and scalability as critical requirements for research process automation, we architected Globus automation services to be cloud-hosted, 
exploiting reliable and scalable cloud services wherever
possible.
We find that these services can scale to support many concurrent clients. 
Experiments show that our 14 initial action providers
exhibit moderate latencies that have proven satisfactory for initial use cases. 
Our experiences applying Globus automation services in several domains has shown that they 
indeed satisfy diverse requirements of varied
research automation use cases. 
Usage has grown rapidly in a short period of time. 

In future work we aim to expand Globus automation services capabilities by integrating a broader set of actions.
To this end, we are exploring methods for transforming various invocation request patterns dynamically and automatically so as to support existing APIs without requiring modification.
We are also exploring methods to decrease action invocation overheads in order to enable lower-latency response times to events and support real-time flow execution.
To improve the flow development process we are actively working on both Pythonic and graphical user interfaces to compose flows.
Continued evaluation of both the platform and user experiences will surely suggest other directions for both research and development.

\section*{Acknowledgments}
We acknowledge the many contributions of the late Steve Tuecke to the design of Globus automation services.
We thank Joe Bottigliero, Jacob Lewis, Uriel Mandujano, Ada Nikolaidis, Rudyard Richter, Stephen Rosen, Seren Thompson, Lei Wang, and others on the Globus product team, and also Nick Saint, Rafael Vescovi, Suresh Narayanan, and Nicholas Schwarz for their support.
This work was supported in part by NSF grant OAC-1835890 and by the U.S.\ Department of Energy under contract DE-AC02-06CH11357.



\bibliographystyle{elsarticle-num}

\balance

\bibliography{truncated_refs}

\begin{thebibliography}{100}
\expandafter\ifx\csname url\endcsname\relax
  \def\url#1{\texttt{#1}}\fi
\expandafter\ifx\csname urlprefix\endcsname\relax\def\urlprefix{URL }\fi
\expandafter\ifx\csname href\endcsname\relax
  \def\href#1#2{#2} \def\path#1{#1}\fi

\bibitem{stach2021autonomous}
E.~Stach, et~al., Autonomous experimentation systems for materials development:
  A community perspective, Matter 4~(9) (2021) 2702--2726.

\bibitem{LEONG20223124}
C.~J. Leong, et~al.,
  \href{https://www.sciencedirect.com/science/article/pii/S259023852200474X}{An
  object-oriented framework to enable workflow evolution across materials
  acceleration platforms}, Matter 5~(10) (2022) 3124--3134.
\newblock \href {https://doi.org/https://doi.org/10.1016/j.matt.2022.08.017}
  {\path{doi:https://doi.org/10.1016/j.matt.2022.08.017}}.
\newline\urlprefix\url{https://www.sciencedirect.com/science/article/pii/S259023852200474X}

\bibitem{liu2021bridging}
Z.~Liu, et~al., Bridging data center {AI} systems with edge computing for
  actionable information retrieval, in: 3rd Annual Workshop on Extreme-scale
  Experiment-in-the-Loop Computing, IEEE, 2021, pp. 15--23.

\bibitem{trifan2022intelligent}
A.~Trifan, et~al., Intelligent resolution: Integrating {Cryo-EM with AI}-driven
  multi-resolution simulations to observe the severe acute respiratory syndrome
  coronavirus-2 replication-transcription machinery in action, The
  International Journal of High Performance Computing Applications (2022)
  10943420221113513.

\bibitem{bpel}
C.~Barreto, et~al., Web {Services Business Process Execution Language Version
  2.0} primer, oASIS Specification (2007).

\bibitem{conductor}
Conductor scalable workflow orchestration, \url{https://conductor.netflix.com}.
  Accessed November 2022.

\bibitem{xin2018developers}
D.~Xin, et~al., How developers iterate on machine learning workflows, in: IDEA
  Workshop at KDD, 2018.

\bibitem{chard14efficient}
K.~{Chard}, et~al., Efficient and secure transfer, synchronization, and sharing
  of big data, IEEE Cloud Computing 1~(3) (2014) 46--55.

\bibitem{amazonstateslanguage}
{Amazon States Language}, \url{https://states-language.net/}. Accessed January
  2022.

\bibitem{GlobusAuth}
S.~{Tuecke}, et~al., Globus {A}uth: A research identity and access management
  platform, in: 12th IEEE International Conference on e-Science, 2016, pp.
  203--212.

\bibitem{ananthakrishnan2015globus}
R.~Ananthakrishnan, et~al., Globus platform-as-a-service for collaborative
  science applications, Concurrency and Computation: Practice and Experience
  27~(2) (2015) 290--305.

\bibitem{vescovi2022linking}
R.~Vescovi, et~al., Linking scientific instruments and computation: Patterns,
  technologies, and experiences, Patterns 3~(10) (2022) 100606.

\bibitem{bicer2021high}
T.~Bicer, et~al., High-performance ptychographic reconstruction with federated
  facilities, in: Smoky Mountains Computational Sciences and Engineering
  Conference, Springer, 2021, pp. 173--189,
  \url{https://arxiv.org/abs/2111.11330}.

\bibitem{blaiszik2019data}
B.~Blaiszik, et~al., A data ecosystem to support machine learning in materials
  science, MRS Communications 9~(4) (2019) 1125--1133.

\bibitem{charbonneau2021making}
A.~L. Charbonneau, et~al., Making {Common Fund} data more findable: Catalyzing
  a data ecosystem, bioRxiv (2021).
\newblock \href {https://doi.org/10.1101/2021.11.05.467504}
  {\path{doi:10.1101/2021.11.05.467504}}.

\bibitem{joachimiak2022new}
D.~A. Sherrell, et~al., Fixed-target serial crystallography at {Structural
  Biology Center}, bioRxiv (2022).
\newblock \href {https://doi.org/10.1101/2022.04.06.487333}
  {\path{doi:10.1101/2022.04.06.487333}}.

\bibitem{levental2022ultrafast}
M.~Levental, et~al., Ultrafast focus detection for automated microscopy, in:
  International Conference on Computational Science, Springer, 2022, pp.
  403--416.

\bibitem{ali2022fairdms}
A.~Ali, et~al., fair{DMS}: Rapid model training by data and model reuse,
  \url{https://arxiv.org/abs/2204.09805} (2022).

\bibitem{diederichs2017serial}
K.~Diederichs, et~al., Serial synchrotron {X}-ray crystallography ({SSX}), in:
  Protein Crystallography, Springer, 2017, pp. 239--272.

\bibitem{dials}
G.~Winter, et~al., {DIALS}: Implementation and evaluation of a new integration
  package, Acta Crystallographica Section D 74~(2) (2018) 85--97.
\newblock \href {https://doi.org/10.1107/S2059798317017235}
  {\path{doi:10.1107/S2059798317017235}}.

\bibitem{uervirojnangkoorn2015enabling}
M.~Uervirojnangkoorn, et~al., Enabling x-ray free electron laser
  crystallography for challenging biological systems from a limited number of
  crystals, Elife 4 (2015) e05421.

\bibitem{hidayetoglu2021memxct}
M.~Hidayetoglu, et~al., Mem{XCT}: Design, optimization, scaling, and
  reproducibility of x-ray tomography imaging, IEEE Transactions on Parallel
  and Distributed Systems 33~(9) (2021) 2014--2031.

\bibitem{liu2020tomogan}
Z.~Liu, et~al., {TomoGAN}: Low-dose synchrotron x-ray tomography with
  generative adversarial networks, JOSA A 37~(3) (2020) 422--434.

\bibitem{lehmkuhler2021femtoseconds}
F.~Lehmk{\"u}hler, et~al., From femtoseconds to hours--measuring dynamics over
  18 orders of magnitude with coherent x-rays, Applied Sciences 11~(13) (2021)
  6179.

\bibitem{maiden2011superresolution}
A.~M. Maiden, et~al., Superresolution imaging via ptychography, JOSA A 28~(4)
  (2011) 604--612.

\bibitem{pokharel18hedm}
R.~Pokharel, Overview of high-energy x-ray diffraction microscopy ({HEDM}) for
  mesoscale material characterization in three-dimensions, in: Materials
  Discovery and Design, Springer International Publishing, 2018, pp. 167--201.
\newblock \href {https://doi.org/10.1007/978-3-319-99465-9_7}
  {\path{doi:10.1007/978-3-319-99465-9_7}}.

\bibitem{dubochet2012cryo}
J.~Dubochet, Cryo-{EM}—the first thirty years, Journal of Microscopy 245~(3)
  (2012) 221--224.

\bibitem{huerta2019enabling}
E.~A. Huerta, et~al., Enabling real-time multi-messenger astrophysics
  discoveries with deep learning, Nature Reviews Physics 1~(10) (2019)
  600--608.

\bibitem{bernier2011far}
J.~V. Bernier, et~al., Far-field high-energy diffraction microscopy: A tool for
  intergranular orientation and strain analysis, The Journal of Strain Analysis
  for Engineering Design 46~(7) (2011) 527--547.

\bibitem{MIDAS}
{MIDAS, Microstructural Imaging using Diffraction Analysis Software},
  \url{https://www.aps.anl.gov/Science/Scientific-Software/MIDAS}. Accessed
  March 2022.

\bibitem{blaiszik16mdf}
B.~Blaiszik, et~al., The {Materials Data Facility}: Data services to advance
  materials science research, {JOM} 68~(8) (2016) 2045--2052.
\newblock \href {https://doi.org/10.1007/s11837-016-2001-3}
  {\path{doi:10.1007/s11837-016-2001-3}}.

\bibitem{chard19dlhub}
R.~Chard, et~al., {DLHub}: Model and data serving for science, in: 33rd IEEE
  International Parallel and Distributed Processing Symposium, 2019, pp.
  283--292.

\bibitem{cfde}
{Common Fund Data Ecosystem (CFDE)},
  \url{https://commonfund.nih.gov/dataecosystem}.

\bibitem{allcock2019petrel}
W.~E. Allcock, et~al., Petrel: A programmatically accessible research data
  service, in: Practice and Experience in Advanced Research Computing, ACM,
  2019, pp. 1--7.

\bibitem{senior2020alphafold}
A.~W. Senior, et~al., Improved protein structure prediction using potentials
  from deep learning, Nature 577~(7792) (2020) 706--710.
\newblock \href {https://doi.org/10.1038/s41586-019-1923-7}
  {\path{doi:10.1038/s41586-019-1923-7}}.

\bibitem{chard16nexus}
K.~Chard, et~al., Globus {N}exus: A platform-as-a-service provider of research
  identity, profile, and group management, Future Generation Computer Systems
  56 (2016) 571--583.
\newblock \href {https://doi.org/10.1016/j.future.2015.09.006}
  {\path{doi:10.1016/j.future.2015.09.006}}.

\bibitem{allen12software}
B.~Allen, et~al., Software as a service for data scientists, Communications of
  the ACM 55~(2) (2012) 81–88.
\newblock \href {https://doi.org/10.1145/2076450.2076468}
  {\path{doi:10.1145/2076450.2076468}}.

\bibitem{ananthakrishnan18platform}
R.~Ananthakrishnan, et~al., Globus platform services for data publication, in:
  Practice and Experience on Advanced Research Computing, PEARC '18, ACM, New
  York, NY, USA, 2018, pp. 14:1--14:7.

\bibitem{ananthakrishnan20identifiers}
R.~Ananthakrishnan, et~al., An open ecosystem for pervasive use of persistent
  identifiers, in: Practice and Experience in Advanced Research Computing, ACM,
  2020, p. 99–105.
\newblock \href {https://doi.org/10.1145/3311790.3396660}
  {\path{doi:10.1145/3311790.3396660}}.

\bibitem{chard2020funcx}
R.~Chard, Y.~Babuji, Z.~Li, T.~Skluzacek, A.~Woodard, B.~Blaiszik, I.~Foster,
  K.~Chard, Func{X}: A federated function serving fabric for science, in: 29th
  International Symposium on High-performance Parallel and Distributed
  Computing, 2020, pp. 65--76.

\bibitem{li2022funcx}
Z.~Li, R.~Chard, Y.~Babuji, B.~Galewsky, T.~J. Skluzacek, K.~Nagaitsev,
  A.~Woodard, B.~Blaiszik, J.~Bryan, D.~S. Katz, I.~Foster, K.~Chard, Func{X}:
  Federated function as a service for science, IEEE Transactions on Parallel
  and Distributed Systems 33~(12) (2022) 4948--4963.

\bibitem{alt20oauthssh}
J.~Alt, et~al., {OAuth SSH with Globus Auth}, in: Practice and Experience in
  Advanced Research Computing, ACM, 2020, pp. 34--40.
\newblock \href {https://doi.org/10.1145/3311790.3396658}
  {\path{doi:10.1145/3311790.3396658}}.

\bibitem{globusactionproviderlist}
{Globus Action Providers},
  \url{https://globus-automate-client.readthedocs.io/en/latest/globus_action_providers.html}.
  Accessed August 2022.

\bibitem{amazonstepfunctions}
{AWS Step Functions Visual workflows for modern applications},
  \url{https://aws.amazon.com/step-functions}. Accessed January 2022.

\bibitem{bhutton-json-schema-00}
A.~Wright, et~al., {JSON Schema}: A media type for describing {JSON} documents,
  Internet-Draft draft-bhutton-json-schema-00, Internet Engineering Task Force,
  work in Progress (Dec. 2020).

\bibitem{oauth2}
D.~Hardt, {OAuth} 2.0 authorization framework specification,
  \url{http://tools.ietf.org/html/rfc6749} (2012).

\bibitem{actionprovidertoolkit}
{Globus Action Provider Tools},
  \url{https://action-provider-tools.readthedocs.io/}. Accessed August 2022.

\bibitem{workflowlist}
Existing workflow systems, \url{
  https://s.apache.org/existing-workflow-systems}. Accessed January 2022.

\bibitem{ludascher2006scientific}
B.~Lud{\"a}scher, et~al., Scientific workflow management and the {K}epler
  system, Concurrency and Computation: Practice and Experience 18~(10) (2006)
  1039--1065.

\bibitem{goecks2010galaxy}
J.~Goecks, et~al., Galaxy: A comprehensive approach for supporting accessible,
  reproducible, and transparent computational research in the life sciences,
  Genome Biology 11~(8) (2010) 1--13.

\bibitem{deelman2015pegasus}
E.~Deelman, et~al., Pegasus, a workflow management system for science
  automation, Future Generation Computer Systems 46 (2015) 17--35.

\bibitem{albrecht2012makeflow}
M.~Albrecht, et~al., Makeflow: A portable abstraction for data intensive
  computing on clusters, clouds, and grids, in: 1st ACM SIGMOD Workshop on
  Scalable Workflow Execution Engines and Technologies, 2012, pp. 1--13.

\bibitem{babuji19parsl}
Y.~Babuji, et~al., Parsl: Pervasive parallel programming in python, in: 28th
  International Symposium on High-Performance Parallel and Distributed
  Computing, ACM, 2019, pp. 25--36.

\bibitem{silva21workflows}
R.~F. da~Silva, et~al., A community roadmap for scientific workflows research
  and development, in: 2021 IEEE Workshop on Workflows in Support of
  Large-Scale Science (WORKS), 2021, pp. 81--90.
\newblock \href {https://doi.org/10.1109/WORKS54523.2021.00016}
  {\path{doi:10.1109/WORKS54523.2021.00016}}.

\bibitem{liew2016scientific}
C.~S. Liew, et~al., Scientific workflows: Moving across paradigms, ACM
  Computing Surveys 49~(4) (2016) 1--39.

\bibitem{krauter02gridmanagement}
K.~Krauter, et~al., A taxonomy and survey of grid resource management systems
  for distributed computing, Software: Practice and Experience 32~(2) (2002)
  135--164.

\bibitem{deelman2009workflows}
E.~Deelman, et~al., Workflows and e-{S}cience: An overview of workflow system
  features and capabilities, Future Generation Computer Systems 25~(5) (2009)
  528--540.

\bibitem{swift}
M.~Wilde, et~al., Swift: A language for distributed parallel scripting,
  Parallel Computing 37~(9) (2011) 633--652.

\bibitem{hull2006taverna}
D.~Hull, et~al., Taverna: A tool for building and running workflows of
  services, Nucleic Acids Research 34~(suppl\_2) (2006) W729--W732.

\bibitem{curbera2002unraveling}
F.~Curbera, et~al., Unraveling the {Web} services web: An introduction to
  {SOAP, WSDL, and UDDI}, IEEE Internet computing 6~(2) (2002) 86--93.

\bibitem{alshuqayran2016systematic}
N.~Alshuqayran, et~al., A systematic mapping study in microservice
  architecture, in: IEEE 9th International Conference on Service-Oriented
  Computing and Applications, IEEE, 2016, pp. 44--51.

\bibitem{candela2021workflow}
L.~Candela, et~al., A workflow language for research e-infrastructures,
  International Journal of Data Science and Analytics 11~(4) (2021) 361--376.

\bibitem{dagman}
{DAGman: The Directed Acyclic Graph Manager},
  \url{http://www.cs.wisc.edu/condor/dagman}.

\bibitem{cwl}
Common workflow language specifications, v1.0.2,
  \url{https://www.commonwl.org/v1.0/}. Accessed April 2020.

\bibitem{emmerich2005grid}
W.~Emmerich, et~al., Grid service orchestration using the business process
  execution language ({BPEL}), Journal of Grid Computing 3~(3) (2005) 283--304.

\bibitem{tan2010comparison}
W.~Tan, et~al., A comparison of using {T}averna and {BPEL} in building
  scientific workflows: the case of {caGrid}, Concurrency and Computation:
  Practice and Experience 22~(9) (2010) 1098--1117.

\bibitem{tan07bpel4job}
W.~Tan, et~al., {BPEL4Job}: A fault-handling design for job flow management,
  in: B.~J. Kr{\"a}mer, K.-J. Lin, P.~Narasimhan (Eds.), International
  Conference on Service-Oriented Computing, Springer Berlin Heidelberg, Berlin,
  Heidelberg, 2007, pp. 27--42.

\bibitem{simpleworkflowservice}
{Amazon Simple Workflow Service},
  \url{https://docs.aws.amazon.com/amazonswf/latest/developerguide/swf-welcome.html}.
  Accessed January 2022.

\bibitem{githubactions}
{GitHub Actions}, \url{https://github.com/features/actions/}. Accessed January
  2022.

\bibitem{codepipeline}
{AWS CodePipeline}, \url{https://aws.amazon.com/codepipeline/}. Accessed
  January 2022.

\bibitem{eugster2003many}
P.~T. Eugster, et~al., The many faces of publish/subscribe, ACM Computing
  Surveys 35~(2) (2003) 114--131.

\bibitem{alqaoud2009publish}
A.~Alqaoud, et~al., Publish/subscribe as a model for scientific workflow
  interoperability, in: 4th Workshop on Workflows in Support of Large-Scale
  Science, 2009, pp. 1--10.

\bibitem{kamburugamuve2015framework}
S.~Kamburugamuve, et~al., A framework for real time processing of sensor data
  in the cloud, Journal of Sensors 2015 (2015).

\bibitem{renart2017online}
E.~Renart, et~al., Online decision-making using edge resources for
  content-driven stream processing, in: 13th International Conference on
  e-Science, IEEE, 2017, pp. 384--392.

\bibitem{epics}
{Experimental Physics and Industrial Control System (EPICS)},
  \url{https://epics.anl.gov}. Accessed August 2022.

\bibitem{quigley2009ros}
M.~Quigley, et~al., {ROS}: An open-source {Robot Operating System}, in: ICRA
  Workshop on Open Source Software, Vol.~3, Kobe, Japan, 2009, p.~5.

\bibitem{xu2017irods}
H.~Xu, et~al., {iRODS primer 2}: Integrated {Rule-Oriented Data System},
  Synthesis Lectures on Information Concepts, Retrieval, and Services 9~(3)
  (2017) 1--131.

\bibitem{ur2014practical}
B.~Ur, et~al., Practical trigger-action programming in the smart home, in:
  Conference on Human Factors in Computing Systems, 2014, pp. 803--812.

\bibitem{ur2016trigger}
B.~Ur, et~al., Trigger-action programming in the wild: An analysis of 200,000
  {IFTTT} recipes, in: Conference on Human Factors in Computing Systems, 2016,
  pp. 3227--3231.

\bibitem{ryanref}
R.~Chard, et~al., High-throughput neuroanatomy and trigger-action programming:
  A case study in research automation, in: 1st International Workshop on
  Autonomous Infrastructure for Science, 2018, pp. 1--7.

\bibitem{goscinski2014multi}
W.~J. Goscinski, et~al., The {Multi-modal Australian ScienceS Imaging and
  Visualization Environment (MASSIVE)} high performance computing
  infrastructure: applications in neuroscience and neuroinformatics research,
  Frontiers in Neuroinformatics 8 (2014) 30.

\bibitem{plale2006casa}
B.~Plale, et~al., {CASA} and {LEAD}: Adaptive cyberinfrastructure for real-time
  multiscale weather forecasting, Computer 39~(11) (2006) 56--64.

\bibitem{elias2017s}
A.~R. Elias, et~al., Where's the bear?--{A}utomating wildlife image processing
  using {IoT} and edge cloud systems, in: IEEE/ACM Second International
  Conference on Internet-of-Things Design and Implementation, IEEE, 2017, pp.
  247--258.

\bibitem{beckman2007spruce}
P.~Beckman, et~al., {SPRUCE}: A system for supporting urgent high-performance
  computing, in: Grid-based Problem Solving Environments, Springer, 2007, pp.
  295--311.

\bibitem{altintas2020using}
I.~Altintas, Using dynamic data driven cyberinfrastructure for next generation
  disaster intelligence, in: International Conference on Dynamic Data Driven
  Application Systems, Springer, 2020, pp. 18--21.

\bibitem{boccali2021dynamic}
T.~Boccali, et~al., Dynamic distribution of high-rate data processing from
  {CERN} to remote {HPC} data centers, Computing and Software for Big Science
  5~(1) (2021) 1--13.

\bibitem{wilkins2008teragrid}
N.~Wilkins-Diehr, et~al., Tera{G}rid science gateways and their impact on
  science, Computer 41~(11) (2008) 32--41.

\bibitem{blaschke2021real}
J.~P. Blaschke, et~al., Real-time {XFEL} data analysis at {SLAC} and {NERSC}: A
  trial run of nascent exascale experimental data analysis, arXiv:2106.11469
  (2021).

\bibitem{cholia2010newt}
S.~Cholia, et~al., {NEWT: A RESTful} service for building high performance
  computing web applications, in: Gateway Computing Environments Workshop,
  IEEE, 2010, pp. 1--11.

\bibitem{stubbs2021tapis}
J.~Stubbs, et~al., Tapis: An {API} platform for reproducible, distributed
  computational research, in: Future of Information and Communication
  Conference, Springer, 2021, pp. 878--900.

\bibitem{thain2005distributed}
D.~Thain, et~al., Distributed computing in practice: The {C}ondor experience,
  Concurrency and Computation: Practice and Experience 17~(2-4) (2005)
  323--356.

\bibitem{salim2019balsam}
M.~Salim, et~al., Balsam: Near real-time experimental data analysis on
  supercomputers, in: 1st IEEE/ACM Annual Workshop on Large-scale
  Experiment-in-the-Loop Computing, IEEE, 2019, pp. 26--31.

\bibitem{nickolay2021towards}
S.~Nickolay, et~al., Towards accommodating real-time jobs on {HPC} platforms,
  \url{https://arxiv.org/abs/2103.13130} (2021).

\bibitem{antypas21enabling}
K.~B. Antypas, et~al., Enabling discovery data science through cross-facility
  workflows, in: 2021 IEEE International Conference on Big Data (Big Data),
  2021, pp. 3671--3680.
\newblock \href {https://doi.org/10.1109/BigData52589.2021.9671421}
  {\path{doi:10.1109/BigData52589.2021.9671421}}.

\bibitem{bard2022lbnl}
D.~Bard, et~al., The {LBNL} superfacility project report (2022).
\newblock \href {https://doi.org/10.48550/arXiv.2206.11992}
  {\path{doi:10.48550/arXiv.2206.11992}}.

\bibitem{bard21superfacility}
D.~J. Bard, et~al., Automation for data-driven research with the {NERSC}
  superfacility {API}, in: H.~Jagode, H.~Anzt, H.~Ltaief, P.~Luszczek (Eds.),
  High Performance Computing, Springer International Publishing, Cham, 2021,
  pp. 333--345.

\bibitem{stansberry2019datafed}
D.~Stansberry, et~al., Data{F}ed: towards reproducible research via federated
  data management, in: International Conference on Computational Science and
  Computational Intelligence, IEEE, 2019, pp. 1312--1317.

\bibitem{sparkes2010towards}
A.~Sparkes, et~al., Towards robot scientists for autonomous scientific
  discovery, Automated Experimentation 2~(1) (2010) 1--11.

\bibitem{roch2018chemos}
L.~M. Roch, et~al., Chem{OS}: orchestrating autonomous experimentation, Science
  Robotics 3~(19) (2018) eaat5559.

\bibitem{steiner2019organic}
S.~Steiner, et~al., Organic synthesis in a modular robotic system driven by a
  chemical programming language, Science 363~(6423) (2019).

\bibitem{burger2020mobile}
B.~Burger, et~al., A mobile robotic chemist, Nature 583~(7815) (2020) 237--241.

\bibitem{noack2021gaussian}
M.~M. Noack, et~al., Gaussian processes for autonomous data acquisition at
  large-scale synchrotron and neutron facilities, Nature Reviews Physics 3~(10)
  (2021) 685--697.

\end{thebibliography}

\end{document}